\documentclass[lettersize,journal]{IEEEtran}
\usepackage{amsmath,amsfonts}
\usepackage{algorithmic}
\usepackage{algorithm}
\usepackage{array}
\usepackage{textcomp}
\usepackage{stfloats}
\usepackage{url}
\usepackage{verbatim}
\usepackage{graphicx}
\usepackage{cite}
\usepackage[caption=false,font=footnotesize]{subfig}

\usepackage{amsmath,amssymb,amsfonts}
\usepackage{xcolor}
\usepackage[normalem]{ulem} 
\usepackage{cancel}
\usepackage{comment}
\usepackage{caption}

\usepackage{multirow}
\usepackage{mathptmx} 
\usepackage{siunitx}
\usepackage{array}
\usepackage{booktabs}
\usepackage[table]{xcolor}
\usepackage{lipsum}
\usepackage{arydshln}
\usepackage[table]{xcolor}
\usepackage{colortbl}
\usepackage{threeparttable}

\usepackage{booktabs}
\usepackage{makecell}
\usepackage{multirow}
\usepackage{rotating}
\usepackage{amsmath}
\usepackage{graphicx}
\usepackage{makecell}
\usepackage{bm}
\usepackage[caption=false,font=footnotesize]{subfig}

\usepackage{tikz}
\usetikzlibrary{arrows.meta,positioning,calc}

\tikzset{
  block/.style={
    draw, rectangle, thick, rounded corners=1mm,
    minimum width=70mm,
    minimum height=12mm,
    align=left,
    font=\small,
    inner sep=3mm
  },
  smallblock/.style={
    draw, rectangle, thick, rounded corners=1mm,
    minimum width=20mm,
    minimum height=9mm,
    align=center,
    font=\small
  },
  titlebox/.style={
    draw, rectangle, thick, rounded corners=1mm,
    minimum width=90mm,
    minimum height=10mm,
    align=center,
    font=\bfseries\small
  },
  line/.style={-Stealth, thick}
}

\usepackage{tikz}
\usetikzlibrary{arrows.meta,positioning,calc}
\newcommand{\redcircle}[1]{%
  \tikz[baseline=(char.base)]{
    \node[
      shape=circle,
      draw=red, fill=white,
      text=red,
      inner sep=.45pt
    ] (char) {#1};
  }%
}

\begin{document}

\title{A Generic Modulo-$(2^n\pm\delta)$ RNS Multiplier \\Based on Twit Representation


}

\author{Saeid Gorgin, 
Amirhossein	Sadr, Behzad Salami, and Dara	Rahmati        

\thanks{Saeid Gorgin is a Lecturer with the School of Physics, Engineering and Computer
Science, University of Hertfordshire, Hatfield, U.K. (s.gorgin@herts.ac.uk).}
\thanks{Amirhossein Sadr is a Researcher with the School of Computer Science, Institute for Research in Fundamental Sciences (IPM), Tehran, Iran (a.sadr@ipm.ir).}
\thanks{Behzad Salami is a Leading Researcher and head of the FPGATech Group at Barcelona Supercomputing Center (BSC) , Barcelona, Spain (behzad.salami@bsc.es).}

\thanks{Dara Rahmati is an Assistant Professor with Department of Computer Science and Engineering at Shahid Beheshti University (SBU), Tehran, Iran (d\_rahmati@sbu.ac.ir).}
}


\maketitle
\begin{abstract}
Modular multiplication is a fundamental arithmetic primitive in Residue Number Systems (RNS) and is often the dominant source of delay, area, and energy consumption in RNS datapaths used in cryptography, signal processing, and machine-learning accelerators. Recent work introduced a twit-based residue representation for moduli of the form $2^n \pm \delta$, with $0 \le \delta \le 2^{n-1}-1$, and showed that it enables efficient generic modular addition and subtraction across the full admissible $\delta$ range. However, an efficient modular multiplier compatible with the same representation has remained unavailable.
This paper presents a generic twit-based modulo-$(2^n \pm \delta)$ multiplier for RNS channels. The proposed architecture computes the product through operand splitting, modular partial-product generation, carry-save accumulation, overflow folding, and a twit-compatible final modular addition. By deferring carry propagation to the final stage, the resulting organization avoids the long critical paths characteristic of conventional multiply-then-reduce designs.
To demonstrate the effectiveness of the proposed approach, we study a modulus set with 5-bit residue channels and show that, owing to the broad admissible range of $\delta$, it can provide a sufficiently wide dynamic range. Moreover, additional 8-bit and 11-bit configurations are used to evaluate the proposed approach at larger channel widths. We implement and synthesize the proposed multiplier in a FreePDK 45\,nm flow, and the results show average reductions of 20.5\% in delay, 13.2\% in area, and 28.0\% in power relative to baseline designs. A system-level study further indicates that these circuit-level improvements translate into lower end-to-end latency over a broad range of modular multiplication and addition workloads.
\end{abstract}


\begin{IEEEkeywords}
Residue Number Systems, Modular Multiplication, Two-Valued Digit Representation,  Finite Field Arithmetic.
\vspace{-10pt}
\end{IEEEkeywords}

\section{Introduction}
\label{sec:intro}

\IEEEPARstart{M}{odular} multiplication is a fundamental arithmetic operation in Residue Number System (RNS) computation and constitutes a major performance and energy bottleneck in a wide range of applications, including cryptography~\cite{ahmadpour2021up}, digital signal processing~\cite{nagornov2022rns}, and neural network acceleration~\cite{valueva2020application}. In such systems, a large portion of the arithmetic workload is spent on repeated modular multiplications, often in the form of multiply--accumulate (MAC) operations, making the efficiency of modular multipliers a critical design concern~\cite{samimi2019res}.

In RNS-based systems, increasing the number of moduli plays a key role from two complementary perspectives. First, employing a larger set of moduli directly increases the achievable dynamic range, which is essential for high-precision computation. Second, distributing the dynamic range across more residue channels enables the use of smaller moduli, reducing the bit-width of each arithmetic unit and allowing higher operating frequencies, lower power consumption, and improved scalability~\cite{parhami2010}. Consequently, the availability of arithmetic units that can efficiently support broad families of moduli is highly desirable.

As in modular addition, where a large number of efficient designs have been proposed for a few highly structured moduli (e.g., $2^n \pm1$), modular multiplication has also been studied much more extensively for special moduli than for broad modulus families. In particular, a rich body of work has addressed multiplier design for special cases such as $2^n-1$, $2^n+1$, and $2^n \pm 3$~\cite{zimmermann1999efficient,efstathiou2005efficient,vergos2007design,efstathiou2011design,efstathiou2014efficient,matutino2010arithmetic,ahmadifar2013improved,jaberipur2023modulo}. In contrast, for the broader family of moduli of the form $2^n \pm \delta$, RNS-oriented hardware solutions remain scarce. To the best of our knowledge, the most relevant prior works in this direction are~\cite{hiasat2000new,matutino2012rns}. Both approaches rely on conventional binary multiplication followed by modular reduction, which results in nontrivial hardware complexity and long carry-propagation paths.

In our prior work~\cite{gorgin2025generic}, we introduced a twit-based residue representation for moduli of the form $2^n \pm \delta$. In this representation, each residue is encoded as an $n$-bit unsigned value augmented by a two-valued digit (twit) that represents either $0$ or $\pm\delta$. We showed that this representation enables efficient generic modular addition and subtraction across the full range
$0 \le \delta \le 2^{n-1}-1$,
by embedding the $\pm\delta$ correction directly into the arithmetic datapath. In the present paper, the term \emph{generic} is used in this precise sense, namely, support for the full admissible $\delta$ range within the modulus family $2^n \pm \delta$, rather than support for arbitrary unrelated moduli. Building on that earlier result, we now address the remaining open problem of designing an efficient modular multiplier compatible with the same twit-based representation.

At the architectural level, rather than forming a full binary product and reducing it afterward, the proposed multiplier begins by splitting each operand into small groups and generating the corresponding weighted partial products directly in modular form. These partial products are then combined through carry-save accumulation, while the overflow contribution produced during accumulation is folded back into the active range before a twit-compatible final modular addition is performed. By avoiding repeated full-width carry-propagation stages, the resulting organization is well suited to compact, high-frequency residue channels.

To demonstrate the efficiency of the proposed approach, we first present a representative case study based on a 12-modulus RNS configuration with 5-bit residue channels, whose overall dynamic range exceeds $2^{64}$. This example shows that generic support over the full $\delta$ range can provide a wide dynamic range even with narrow residue channels.

The proposed approach is evaluated through both analytical and experimental methods. The analytical study indicates the expected reduction in critical-path delay, while synthesis results quantify the corresponding improvements in delay, area, and power. Although the $n=5$ case already provides a wide dynamic range sufficient for many applications, additional evaluations are also conducted for larger channel widths, namely $n=8$ and $n=11$. Since the number of admissible offsets grows significantly with $n$, only representative values of $\delta$ are selected for these larger-width cases. 
In addition, a system-level case study is used to assess the impact of the proposed approach in a MAC-dominated accelerator setting. 

The main contributions of this paper are summarized as follows:

\begin{itemize}
    \item A generic twit-based modular multiplication approach for RNS channels with moduli of the form $2^n \pm \delta$, valid over the full range $0 \le \delta \le 2^{n-1}-1$.

    \item A modular multiplication organization based on operand splitting, weighted modular partial-product generation, and carry-save reduction.

    \item An overflow-folding and twit-compatible final addition scheme that avoids conventional full-width multiply-then-reduce datapaths.

    \item A representative implementation on a 12-modulus RNS configuration with 5-bit residue channels and dynamic range exceeding $2^{64}$, together with additional evaluations for larger channel widths.

    \item Analytical and synthesis-based evaluations demonstrating the effectiveness of the proposed approach.
\end{itemize}
\vspace{5pt}
\textbf{\textit{Availability}:} 
To promote reproducibility, transparency, and further research, all HDL codes for the proposed implementations are publicly available at https://github.com/GorginSaeid/Generic-Twit-Based-Multiplier

The remainder of this paper is organized as follows. Section~\ref{sec:background} reviews the fundamental concepts underlying the proposed work, including RNS, modular multiplication based on operand partitioning, and the twit-based residue representation for moduli of the form $2^n \pm \delta$. Section~\ref{sec:prior} reviews related work on modular multiplication. Section~\ref{sec:main} presents the proposed multiplication approach and architecture. Section~\ref{sec:eval} reports the analytical and experimental results. Finally, Section~\ref{sec:conclusion} concludes the paper.


\section{Background}
\label{sec:background}
This section reviews the key concepts that form the foundation of the proposed modular multiplication architecture. First, we briefly revisit the principles of the RNS and highlight the central role of modular multiplication as a dominant computational bottleneck in RNS-based processing. Next, we review the standard principle of constructing large multipliers from smaller multiplier blocks through operand partitioning. Finally, we revisit the twit-based residue representation for moduli of the form $2^n \pm \delta$.

\subsection{Residue Number Systems and Modular Multiplication}
\label{subsec:rns_mult}

The Residue Number System (RNS) is a non-weighted number representation in which an integer
$X \in [0,M)$ is uniquely represented by a set of residues
\begin{align}
X \;\longleftrightarrow\; (x_1, x_2, \dots, x_k), \qquad x_i = X \bmod m_i,
\end{align}
where the moduli $\{m_1,m_2,\dots,m_k\}$ are pairwise coprime and the dynamic range is
$M=\prod_{i=1}^{k} m_i$~\cite{mohan2016}.
A key advantage of RNS lies in its inherent parallelism: arithmetic operations on $X$ and $Y$
can be decomposed into independent operations on each residue channel without carry
propagation across moduli~\cite{parhami2010}.

Among all arithmetic operations in RNS, modular multiplication plays a central role~\cite{bajard2001modular}.
Given two numbers $A$ and $B$ represented in RNS, multiplication is performed independently
in each residue channel as
\begin{align}
z_i = |a_i \cdot b_i|_{m_i}, \qquad 1 \le i \le k.
\end{align}
In contrast to modular addition, modular multiplication is significantly more expensive,
as it involves partial-product generation followed by modular reduction~\cite{parhami2010}.
Consequently, modular multipliers dominate the overall delay, area, and power consumption
in many RNS-based systems, particularly in cryptographic primitives~\cite{ahmadpour2021up}, digital signal processing~\cite{nagornov2022rns},
and neural network accelerators~\cite{valueva2020application}.

Existing RNS multiplier designs typically follow one of two approaches.
The first relies on conventional binary multiplication followed by a reduction modulo $m_i$,
which leads to large hardware overhead and long critical paths~\cite{hiasat2000new, matutino2012rns}.
The second exploits special properties of specific moduli (e.g., $2^n \pm 1$), enabling more efficient reduction schemes~\cite{muralidharan2012area,radhakrishnan1992novel,sousa2005universal}.
However, such designs lack generality and do not scale well when a large and flexible set
of moduli is required to achieve high dynamic range with narrow residue channels.
This limitation motivates the development of \emph{generic} modular multiplication
architectures that can efficiently support broad classes of moduli.

\subsection{Implementing Large Multipliers Using Smaller Ones}
\label{subsec:karatsuba}

Constructing large multipliers from smaller multiplier blocks is a standard technique in computer arithmetic and digital hardware design~\cite{koren2001computer}. A classical way to motivate this principle is by splitting operands, as also seen in Karatsuba-style multiplication~\cite{karatsuba1962multiplication}. For two $n$-bit operands $A$ and $B$, a basic partitioning into high and low parts can be written as
\begin{align}
A = A^H \cdot 2^{n/2} + A^L, \qquad
B = B^H \cdot 2^{n/2} + B^L,
\end{align}
where $A^H$, $A^L$, $B^H$, and $B^L$ denote the high and low halves of the operands. Their product is then expressed as
\begin{align}
A \cdot B
&= (A^H B^H)\cdot 2^n
+ (A^H B^L + A^L B^H)\cdot 2^{n/2}
+ (A^L B^L).
\label{eq:partition_mult}
\end{align}

It is important to clarify that the present work uses only the \emph{operand-splitting principle} of this formulation. Unlike the classical Karatsuba algorithm, which reduces the number of half-width multiplications from four to three by introducing additional additions and subtractions, the proposed architecture does not exploit that optimization. Instead, it preserves all local products and maps them to a regular modular datapath. This choice is more suitable for the present context, where weighted local products must be generated directly in modular form and later accumulated through structured reduction.

Under modular arithmetic, the same splitting idea can be applied by reducing each weighted term modulo $m_i$:
\vspace{-5pt}
\begin{align}
\label{eq:karatsuba_mod}
|A \cdot B|_{m_i}
&= \Big|
|(A^H B^H)\cdot 2^n|_{m_i}
\nonumber 
+ |(A^H B^L + A^L B^H)\cdot 2^{n/2}|_{m_i}
\nonumber\\
&\quad + |(A^L B^L)|_{m_i}
\Big|_{m_i}.
\end{align}

By maintaining all partial products in modular form, the multiplication can be
organized as a sequence of modular additions, avoiding full-width
binary multiplication followed by costly correction logic.
This formulation is particularly advantageous for hardware implementations,
where regularity and predictable data paths are essential for scalability.

\subsection{Two-Valued Digit Representation}
\label{sec:Twit}

A two-valued digit, analogous to a binary digit that assumes one of the two values 0 and 1, represents one of two distinct numerical values, denoted by \(\alpha\) and \(\beta\)~\cite{parhami2010}. In the present work, this concept is used to encode the correction term \(\pm\delta\) through the digit set \(\{0,\pm\delta\}\).

The proposed representation is based on the Weighted Bit-Set (WBS) framework, which generalizes conventional binary encoding by allowing digits to represent weighted value sets rather than only \(\{0,1\}\)~\cite{jaberipur2005}. Within this framework, the two-valued digit adopted here is referred to as a \emph{Twit}. A Twit is a binary variable defined by a lower value \(L\) and a gap \(G\), and therefore represents the set \(\{L,L+G\}\). In this work, we choose \(L=0\) and \(G=\pm\delta\), which yields the digit set \(\{0,\pm\delta\}\). This gap-based encoding provides a compact and hardware-friendly way of embedding modular correction into arithmetic over moduli of the form \(2^n \pm \delta\).

\noindent
\textbf{Example 1:}
Let a Twit be defined by \(L=-12\) and \(G=29\). The two values represented by this Twit are then \(\{-12,17\}\). If the Twit is set to 0, it represents \(-12\), whereas if it is set to 1, it represents \(17\).


\section{Related Work}
\label{sec:prior}
This section first reviews representative designs for special moduli, mainly to identify
recurring optimization principles, and then discusses prior efforts on generic modular
multiplication in greater detail.

\subsection{Modular Multipliers for Special Moduli}

Special forms of moduli have long been favored in RNS implementations because they admit
simplified modular reduction and optimized hardware structures.
Among these, moduli of the form $2^n \pm 1$ and $2^n - 3$ have been studied
extensively.

\subsubsection{Modulo-$(2^n \pm 1)$}

A large body of work has focused on multipliers for modulo-$(2^n-1)$ and modulo-$(2^n+1)$,
as these moduli admit simple end-around-carry reductions~\cite{efstathiou2004fast,sousa2005universal,vergos2007design}.
Early designs combined conventional binary multiplication with cyclic shifts and modular
additions, resulting in memoryless architectures with moderate complexity
~\cite{Hiasatfix,hiasat2000new}.
Subsequent efforts emphasized purely combinational VLSI realizations, reducing latency
by avoiding sequential correction steps
\cite{radhakrishnan1992novel,hiasat1996combinational}.

Later refinements incorporated advanced techniques such as modified Booth recoding,
parallel-prefix adders, and diminished-1 representations, significantly reducing the
number of partial products and accelerating modular reduction
\cite{efstathiou2011design,efstathiou2014efficient}.
Additional optimizations further improved both multiplier and adder structures for
modulo-$(2^n+1)$ arithmetic
\cite{vergos2007design,vergos2006novel}.
Universal architectures were also proposed to maintain efficiency while avoiding
explicit Booth recoding
\cite{sousa2005universal,chaves2005faster}.
Collectively, these works established modulo-$(2^n \pm 1)$ multipliers as highly optimized
building blocks for cryptographic and signal-processing applications.

\subsubsection{Modulo-$(2^n - 3)$}

To expand balanced moduli sets beyond $2^n \pm 1$, several studies have investigated
modulo-$(2^n - 3)$ arithmetic.
Matutino \textit{et al.}~\cite{matutino2010arithmetic} proposed a multiplier architecture
based on residue periodicity, reformulating the product as
$|A \times B|_{(2^n-3)} = |3P_h + P_l|_{(2^n-3)}$.
Although this approach employs carry-save reduction to mitigate delay, it relies on
specialized compressors and requires additional correction steps when excess-modulo
operands occur.

Subsequent work improved latency by retaining intermediate results in carry-save form,
thereby eliminating one carry-propagate stage; however, the complexity associated with
handling excess-modulo values remained
\cite{ahmadifar2013improved}.
Parallel-prefix adders combined with double-residue representations were later introduced
to further accelerate modular operations in this domain
\cite{jaberipur2015parallel}.
More recently, Seidel~\cite{seidel2018high} proposed a unified carry-save framework in which
partial products are reorganized and reduced together, integrating Booth recoding and
double-representation modular adders.

The most recent contributions in this line of research adopt modular partial-product
generation and reduction, resulting in systematic and scalable architectures for
modulo-$(2^n-3)$ multiplication~\cite{jaberipur2023modulo}.
While these designs achieve high efficiency, their applicability remains confined to
specific offsets.

\subsection{Generic Modular Multipliers}
Despite extensive research on multipliers for special moduli, relatively few works have explored \emph{generic} modular multiplication architectures for arbitrary coprime moduli, which are essential in scalable RNS systems to extend dynamic range while preserving narrow residue channels.
Early approaches relied on lookup tables for modular reduction~\cite{hiasat2000new, paliouras2001low, hiasat2002rns}, but their poor scalability in area and delay
motivated a transition toward arithmetic-based designs using adders, shifters, and constant multipliers.

\begin{figure}[t]
\centering
\includegraphics[width=0.95\linewidth]{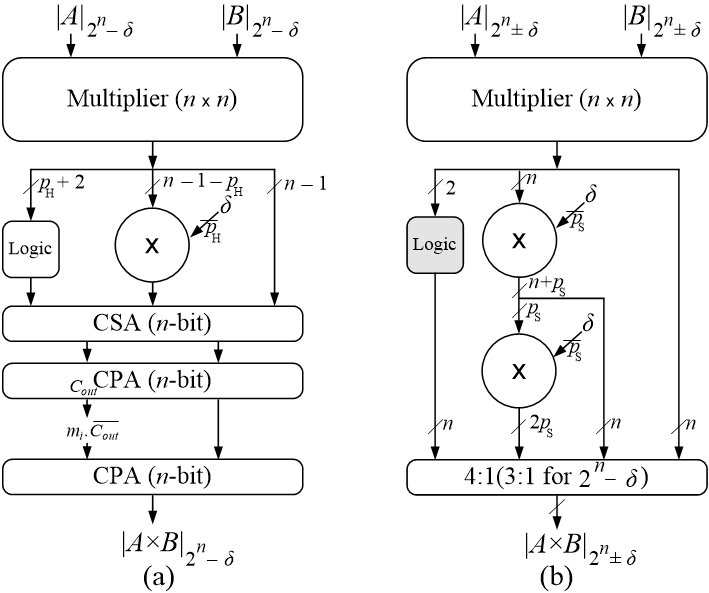}
\caption{Abstract views of generic modular multipliers:
(a) Hiasat~\cite{hiasat2000new} and
(b) Matutino \emph{et al.}~\cite{matutino2012rns} (The gray logic block exists only for moduli of the form \(2^n+\delta\)).
}
\label{fig:Abstract_Hiasat_Sousa}
\vspace{-10pt}
\end{figure}

Representative arithmetic-based generic modular multipliers are shown in
Fig.~\ref{fig:Abstract_Hiasat_Sousa}.
The architecture in Fig.~\ref{fig:Abstract_Hiasat_Sousa}(a), proposed by
Hiasat~\cite{hiasat2000new}, targets moduli of the form $m = 2^n - \delta$ and
decomposes multiplication into high and low partial products followed by reduction
through additions and shifts.
While substantially more scalable than ROM-based designs, it still involves multiple
wide carry-propagation paths whose complexity grows with operand width $n$.

The design in Fig.~\ref{fig:Abstract_Hiasat_Sousa}(b), introduced in~\cite{matutino2012rns},
extends this principle to moduli of the form $2^n \pm \delta$ and demonstrates
near-specialized performance with moderate area overhead.
However, it continues to depend on full-width carry-propagate additions and
nontrivial correction logic.
Moreover, as evident from the structural constraints in Fig.~\ref{fig:Abstract_Hiasat_Sousa}(b),
the supported range of $\delta$ is restricted: specifically, $\delta$ must be
strictly smaller than $2^{\lfloor n/2 \rfloor}$ (i.e., its bit-width, which is shown with $p$ in Fig.~\ref{fig:Abstract_Hiasat_Sousa}(b) is limited to
at most half that of $n$).
This requirement significantly constrains the generality of the design and limits
its applicability to truly arbitrary moduli.

Although existing generic modular multipliers establish the feasibility
of arithmetic-based approaches beyond special moduli, they typically rely on
full-width binary multiplication.
The use of multiple constant multipliers, ROM components, and multi-operand
modular adders increases architectural complexity and hinders scalability,
particularly for large operand sizes and wide modulus sets.
These limitations motivate the development of alternative architectures.

\section{Proposed Generic Modulo-$(2^n \pm \delta)$ Multiplication Framework}
\label{sec:main}
This section presents the proposed generic multiplication framework for moduli of the form $2^n \pm \delta$. Since the multiplier operates directly on twit-encoded operands, we begin by revisiting the operand representation used throughout the architecture. Although this representation has been introduced in our prior work~\cite{gorgin2025generic}, a brief summary is included here for self-containment. We then present a high-level architectural overview, describe the generic algorithm and modular architecture in detail, and finally provide a concrete instantiation for $n=5$ as a representative case study.

\subsection{Operand Representation}
\label{subsec:operandRepresentation}

In the proposed framework, each residue is represented as an $n$-bit unsigned value augmented by a twit. Thus, a residue $A$ is expressed as
\begin{align}
A \;\equiv\; a_{n-1}a_{n-2}\ldots a_1a_0 + \delta_a,
\label{eq:twit_repr}
\end{align}
where $a_{n-1}a_{n-2}\ldots a_1a_0$ is an $n$-bit binary value and
$
\delta_a \in \{0,\pm\delta\}.
$
This scheme is conceptually similar to the double least significant bit (LSB) representation~\cite{parhami2008double}; however, instead of introducing an additional ordinary bit, it uses a twit to encode the $\pm\delta$ correction.

A key advantage of this representation is that all codewords remain valid. For moduli of the form $2^n-\delta$, every residue admits more than one equivalent representation, including forms whose binary part may exceed the modulus or correspond to a negative value prior to modular interpretation. For moduli of the form $2^n+\delta$, only a subset of residues has dual representations. This natural redundancy is useful because it embeds the modular correction directly into the operand representation, rather than requiring explicit compare-and-subtract logic.

The resulting encoding is especially beneficial for modular arithmetic. Since the end-around correction associated with $\pm\delta$ is already captured by the twit, modular addition and subtraction can be implemented with lightweight combinational logic and a single carry-propagate addition~\cite{gorgin2025generic}. Moreover, the same representation provides a unified format for both $2^n-\delta$ and $2^n+\delta$, which is important for generic arithmetic over the full admissible $\delta$ range.

\vspace{5pt}
\noindent
\textbf{Example 2:} Consider $n=5$ and $\delta=5$.

\begin{itemize}
    \item For modulo-$(2^5-5)$, the value $16$ can be represented in two different forms: $10000\mathbf{0}$ and $10101\mathbf{1}$. In the first form, where $\delta_a=\mathbf{0}$, the representation corresponds directly to the binary value $(10000)_2=16$. In the second form, where $\delta_a=\mathbf{1}$, the binary part corresponds to $(10101)_2=21$, and the twit contributes $-5$, so the represented value is again $21-5=16$.

    \item For modulo-$(2^5+5)$, the same value $16$ can be represented as $10000\mathbf{0}$ and $01011\mathbf{1}$. Again, the first form corresponds directly to $(10000)_2=16$. In the second form, the binary part corresponds to $(01011)_2=11$, and the twit contributes $+5$, yielding $11+5=16$.
\end{itemize}
\vspace{5pt}

This example illustrates both the compactness and the redundancy of the twit-based encoding. In the context of multiplication, these properties are particularly important: partial-product generation must correctly account for twit-encoded operands, while the subsequent reduction and final accumulation stages must preserve the benefits of the representation without introducing excessive overhead.

\subsection{Architectural Overview}
\label{subsec:overview}
The proposed generic modulo-\((2^n \pm \delta)\) multiplication framework decomposes a wide-operand
multiplication into four well-defined processing stages, each designed to isolate carry-intensive
operations and maximize modular parallelism. The overall structure of the framework is illustrated in
Fig.~\ref{fig:overall_structure}, which consists of the following four stages:
\begin{itemize}\setlength{\itemsep}{0pt}
    \item \emph{Operand Splitting} 
    \item \emph{Partial Product Generation} 
    \item \emph{Multi-Operand Reduction} 
    \item \emph{Final Modular Addition}
\end{itemize}

The operand splitting stage~\redcircle{1} performs a structural decomposition of the input operands into small digit groups suitable for efficient hardware realization. In the proposed design, the operands are partitioned into groups of three bits. In the most significant groups, where the remaining operand width may be smaller, the group size is reduced accordingly. This stage does not perform arithmetic operations, but rather defines the structural granularity of the subsequent computations.

In the partial product generation stage~\redcircle{2}, the corresponding operand groups are multiplied using combinational logic, producing a set of intermediate partial products. Since each multiplication operates on two three-bit operand segments, the generated partial product can be realized as a six-input Boolean function. This choice keeps the local combinational logic compact and regular, while also being well suited for FPGA implementations, where a six-input function can be efficiently mapped onto a single LUT6. This process can be viewed as a generalized expansion of the multiplication operation used in the underlying operand-decomposition-based multiplication formulation in Eq.~(\ref{eq:karatsuba_mod}). Each partial product is associated with a specific modular channel and serves as an input to the subsequent reduction network.

The reduction stage~\redcircle{3} compresses the generated partial products using multi-operand compressors, such as 3:2 or 4:2 structures, until a depth-two representation is obtained. Owing to this compression process, the bit-width of the intermediate results may exceed the target modulus size \(n\).

The final modular addition stage~\redcircle{4} operates on the reduced operands and produces the result in each modulus. A key design principle of the proposed framework is that partial-product generation, reduction, and final addition are all carried out independently within each modular channel. This modular independence enables parallel instantiation across different moduli and allows the architecture to scale naturally with the desired dynamic range.

It should be noted that, as \(n\) increases, the number of generated partial products also increases, and therefore the reduced outputs of Stage~\redcircle{3} may have a larger bit-width. Since the input size of the combinational logic blocks is restricted to six for regularity and implementation efficiency, an intermediate bit-width reduction operation, referred to here as \emph{squeezing}, may be inserted before Stage~\redcircle{4}, when needed, to reduce the width of the reduced operands. This squeezing operation is treated as an implementation-oriented optimization rather than as part of the core four-stage algorithm, and is explained further in the next subsection.

\subsection{Generic Algorithm and Modular Architecture}
\label{subsec:generic_alg_arch}

This subsection presents the complete algorithmic flow and modular architecture of the proposed generic modulo-$(2^n \pm \delta)$ multiplication framework. Let the target modulus be $m_i = 2^n \pm \delta$, and let $A,B \in [0,m_i-1]$ denote the input operands. The objective is to compute
\begin{align}
P = |A \times B|_{m_i}
\end{align}
through a sequence of structured processing stages that preserve modular independence and avoid long carry-propagation paths. Algorithm~\ref{alg:generic_mult} summarizes the overall flow, and the individual stages are described below.

\subsubsection*{\textnormal{\redcircle{1}} Operand Splitting}
\label{subsubsec:stage1_split}

To enable efficient local product generation while preserving the twit-based residue representation introduced in Section~\ref{subsec:operandRepresentation}, each operand is decomposed into groups of width three. Special care is taken for the least-significant group, which incorporates the twit value. Let a residue $A$ be represented as
\begin{align}
A \equiv \sum_{j=0}^{n-1} a_j 2^j + \delta_a,
\qquad
\delta_a \in \{0,\pm\delta\},
\end{align}
and similarly for operand $B$.

The total number of groups is $\Gamma = 1 + \lceil (n-2)/3 \rceil$.
The least-significant groups are defined as
$
g^{A}_{0}\triangleq (\delta_a, a_1, a_0)$ and $g^{B}_{0}
 \triangleq (\delta_b, b_1, b_0),
$
where each group consists of the twit value together with the two least-significant bits of the binary operand.
The remaining \((\Gamma-1)\) binary bits are partitioned into conventional 3-bit groups starting from bit position 2 ($1 \le \gamma \le \Gamma-1$):
$
g^{A}_{\gamma} \triangleq (a_{3\gamma+1}, a_{3\gamma}, a_{3\gamma-1})$ and $
g^{B}_{\gamma} \triangleq (b_{3\gamma+1}, b_{3\gamma}, b_{3\gamma-1}).
$
If the most-significant group is incomplete, the absent bits are simply omitted.

The group index implicitly determines the positional weight of each group in the original operand. In particular, the least-significant group has zero shift, while the remaining groups are interpreted according to their bit locations in the operand word. Accordingly, the local product block associated with indices $(\gamma,\eta)$ generates the product of the corresponding groups together with the positional weight implied by their indices. This convention allows the weighted contribution of each local product to be handled directly in modular form.

\subsubsection*{\textnormal{\redcircle{2}} Partial Product Generation}
\label{subsubsec:stage2_ppg}

The generation of partial products follows the operand-decomposition principle
introduced in Section~\ref{subsec:karatsuba}, where wide-precision multiplication
is expressed as a weighted combination of smaller modular products.
Using the group-based operand splitting defined in
Section~\ref{subsubsec:stage1_split}, each partial product is generated as
\begin{equation}
PP_{\gamma,\eta} \triangleq |g^{A}_{\gamma} \times g^{B}_{\eta}|_{m_i},
\qquad
0 \le \gamma,\eta \le \Gamma-1,
\label{eq:pp_group}
\end{equation}
where $g^{A}_{\gamma}$ and $g^{B}_{\eta}$ are the $\gamma$-th and $\eta$-th operand groups, respectively, and $m_i$ is the corresponding modulus.

For \(\gamma,\eta \ge 1\), both groups consist of three binary bits.
For \(\gamma=0\) or \(\eta=0\), the corresponding group additionally incorporates the twit
component; however, since the twit represents a constant offset drawn from
\(\{0,\pm\delta\}\), it can be encoded using a single binary control signal.
As a result, each partial product, together with its reduction modulo \(m_i\),
can be realized using a single 6-input Boolean logic function.

\begin{figure}[t]
\centering
\includegraphics[width=0.800\linewidth]{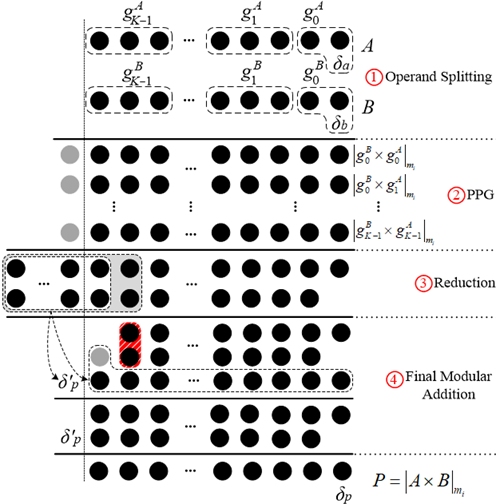}
\vspace{-3pt}
\caption{The overall structure of the proposed multiplier}
\vspace{-5pt}
\label{fig:overall_structure}
\end{figure}

It is worth noting that the raw bit-width of each partial product depends on the modulus.
For $2^n-\delta$, it fits within $n$ bits, whereas for $2^n+\delta$
it may extend to an $(n+1)$-bit width. These additional bits are illustrated by the gray dots in Fig.~\ref{fig:overall_structure}, Stage~\redcircle{2}, at the most-significant position. 
Although this does not affect the
reduction stage, it makes the squeezing step more involved for
$2^n+\delta$, as illustrated in Section~\ref{subsec:case_n5}.

Moreover, since the number of generated partial products is \(\Gamma^2\), the reduction network grows approximately quadratically with \(n\). Consequently, the hardware cost becomes more significant for larger channel widths. Nevertheless, due to the generic nature of the proposed framework and the wide range of admissible \(\delta\) values, even relatively small values of \(n\) can provide a sufficient dynamic range for a broad class of applications.

\subsubsection*{\textnormal{\redcircle{3}} Multi-Operand Reduction}
\label{subsubsec:stage3_reduction}

Since all partial products are generated directly in the modular domain, the subsequent reduction stage performs only accumulation and does not introduce any additional modular correction. For each modular channel \(m_i\), the accumulated value is defined as
\begin{equation}
S_i \;\triangleq\;
\sum_{\gamma=0}^{\Gamma-1}\sum_{\eta=0}^{\Gamma-1}
PP_{\gamma,\eta}.
\label{eq:accumulate_mod_channel}
\end{equation}

In hardware, the summation in Eq.~\eqref{eq:accumulate_mod_channel} is realized using a multi-operand compressor tree. When only 3:2 counters are used, the number of reduction levels is
\(\lambda=\left\lceil \log_{3/2}({\Gamma^2}/2)\right\rceil\), which reduces the large set of modular operands to a depth-two carry-save representation \(\big(S_i^{(0)}, S_i^{(1)}\big)\). In practice, 4:2 compressors can also be employed to reduce the number of reduction levels and achieve a more efficient implementation. Because all operands belong to the same modular channel, the accumulation preserves full modular independence and avoids long carry-propagation paths. The resulting carry-save pair is then forwarded to the final stage.

\subsubsection*{\textnormal{\redcircle{4}} Final Modular Addition}
\label{subsubsec:stage4_final}

After the multi-operand reduction stage, the intermediate result for each modular channel is available in depth-two carry-save form as \(\big(S_i^{(0)}, S_i^{(1)}\big)\). Although all partial products are generated directly modulo \(m_i\), the reduction process itself may increase the bit-width beyond the nominal operand width \(n\).

To obtain the final result, the proposed architecture applies a twit-compatible final addition procedure. In this stage, for moduli of the form \(2^n-\delta\), the bit positions from \(n-1\) upward are first processed by a fixed combinational logic block, shown by the white rectangle in Fig.~\ref{fig:overall_structure}, Stage~\redcircle{3}. This block transforms their aggregate contribution into an equivalent representation consisting of an \(n\)-bit value together with a twit.
For moduli of the form \(2^n+\delta\), the same operation is performed with one additional bit of width, starting from position \(n-2\), as indicated by the gray rectangle in Fig.~\ref{fig:overall_structure}, Stage~\redcircle{3}. In this case, the transformed output is represented in an \((n+1)\)-bit double-MSD format together with a twit.

This transformed contribution is then combined with the lower portion using a carry-save addition, followed by a carry-propagate addition. For \(2^n+\delta\), the bits at position \(n-2\) in Stage~\redcircle{4}, shown by the red-shaded column, are set to zero. If the carry-out of the carry-propagate adder is equal to one, the twit value is corrected accordingly~\cite{gorgin2025generic}, and the final result is obtained.

Accordingly, the final modular result is obtained without requiring repeated full-width carry propagation inside the reduction network. Since this is the only stage in which carry propagation is introduced, the overall critical path remains substantially shorter than in conventional multiply-then-reduce architectures, where full-width carry-propagate additions appear repeatedly throughout the datapath.

\begin{algorithm}[t]
\small
\caption{Generic Modulo-$(2^n \pm \delta)$ Multiplication}
\label{alg:generic_mult}
\begin{algorithmic}[1]
\REQUIRE Operands $A,B \in [0,m_i-1]$, modulus $m_i = 2^n \pm \delta$
\ENSURE $P = |A \times B|_{m_i}$

\vspace{0.3em}
\STATE \emph{Stage}~\redcircle{1} \emph{(Operand Splitting):}
Partition $A$ and $B$ into
$\Gamma = 1 + \lceil (n-2)/3 \rceil$ operand groups
$\{g^{A}_{\gamma}\}$ and $\{g^{B}_{\eta}\}$,
where the least-significant groups incorporate the twit values.

\vspace{0.3em}
\STATE \emph{Stage}~\redcircle{2} \emph{(Partial Product Generation):}
For all $0 \le \gamma,\eta \le \Gamma-1$, generate modular partial products
\[
PP_{\gamma,\eta} \triangleq |g^{A}_{\gamma} \times g^{B}_{\eta}|_{m_i}.
\]

\STATE \emph{Stage}~\redcircle{3} \emph{(Multi-Operand Reduction):}
Accumulate all $PP_{\gamma,\eta}$ belonging to the same modular channel using a
compressor tree, reducing the operand set to a depth-two carry-save
representation  \(\big(S_i^{(0)}, S_i^{(1)}\big)\).

\vspace{0.3em}
\STATE \emph{Stage}~\redcircle{4} \emph{(Final Modular Addition):}
Resolve the carry-save form and compute the final result
\[
P = |S_i^{(0)}+ S_i^{(1)}|_{m_i}.
\]
\vspace{-15pt}

\STATE \emph{return} $P$
\end{algorithmic}
\end{algorithm}

\subsubsection*{Optional Squeezing for Larger Channel Widths}
\label{subsubsec:optional_squeezing}

After Stage~\redcircle{3}, the intermediate result in each modular channel is available in depth-two carry-save form. Although all partial products are generated directly modulo \(m_i\), the reduction process itself may increase the bit-width beyond the nominal operand width \(n\).

Let the integer value represented by the reduced carry-save pair for modulus channel \(m_i\) be denoted by \(S_i\). We decompose it as
\begin{equation}
S_i = S_i^{L} + 2^n S_i^{H},
\label{eq:squeeze_decomp}
\end{equation}
where \(S_i^{L}\) contains the lower \(n\) bit positions, and \(S_i^{H}\) represents the aggregate contribution of all bit positions above \(n-1\).
For moduli of the form \(m_i = 2^n \pm \delta\), the congruence
$
2^n \equiv \mp \delta \pmod{2^n \pm \delta}
$
holds. Therefore, the overflow contribution \(2^n S_i^{H}\) can be replaced by an equivalent modular quantity, and Eq.~\eqref{eq:squeeze_decomp} can be rewritten as
\begin{equation}
S_i \equiv S_i^{L} + \left|2^n S_i^{H}\right|_{m_i}.
\label{eq:squeeze_basic}
\end{equation}

In the proposed implementation, the quantity \(\left|2^n S_i^{H}\right|_{m_i}\) is realized by a fixed combinational logic block. To preserve regularity and implementation efficiency, the number of inputs of this logic block is restricted to at most six. If the overflow part \(S_i^{H}\) contains no more than six bits, the corresponding modular value is generated directly and injected as an additional operand into the carry-save datapath.

When the overflow part exceeds this input limit, squeezing is performed iteratively. In each step, only a bounded subset of the overflow bits is folded back into the active modular range and accumulated through carry-save addition. The procedure continues until the width of the resulting carry-save representation is reduced to at most \(n+2\) bits, thereby making it directly compatible with Stage~\redcircle{4}. Hence, squeezing is introduced only when required, and is regarded as an implementation-oriented refinement for larger channel widths.

\subsection{Case Study: A Representative Moduli Set for \(n=5\) }
\label{subsec:case_n5}
Based on the moduli-set structure \(\{2^{2n},\,2^n \pm \delta\}\), we construct a representative 12-modulus RNS set for the case \(n=5\).
The purpose of this case study is twofold: (\emph{i}) how the framework is specialized for a concrete \(n\), and (\emph{ii}) how a practical set of pairwise-coprime moduli can be formed while retaining implementation-friendly structures.

\subsubsection{Selected Moduli Set}
\label{subsubsec:n5_modset}

For \(n=5\), the base is \(2^n = 32\). We consider the following moduli set:
$\mathcal{M}  = \{17, 19, 23, 29, 31, 1024, 35, 37, 39, 41, 43, 47\}.$ All moduli in \(\mathcal{M}\) are pairwise coprime, enabling parallel multi-channel
processing and straightforward CRT-based reconstruction when needed.

\paragraph*{Mapping to the \(2^n \pm \delta\) form}
Except for the power-of-two channel \(1024\) (i.e., \(2^{2n}\)), the remaining moduli are
centered around \(2^5 = 32\) and can be expressed in the form \(m_i = 2^5 \pm \delta\).
Specifically, the set \(\{17,19,23,29,31\}\) corresponds to
\(\{32-15,\,32-13,\,32-9,\,32-3,\,32-1\}\), while
\(\{35,37,39,41,43,47\}\) corresponds to
\(\{32+3,\,32+5,\,32+7,\,32+9,\,32+11,\,32+15\}\).
As a result, \(\delta \in \{1,3,5,7,9,11,13,15\}\), and for \(n=5\) we have
\(\delta \leq 2^{n-1}-1 = 15\), which is favorable for compact twit-based correction logic.

\paragraph*{Dynamic range}
The effective dynamic range of the set is
$ M \triangleq \prod_{m \in \mathcal{M}} m_i.$
For the selected set $\mathcal{M}$, we obtain
$M = 28{,}620{,}324{,}425{,}937{,}054{,}720 \approx 2^{65}.$
Hence, the constructed multi-channel system provides approximately \(65\) bits of
dynamic range.

\subsubsection{Framework Specialization}
\label{subsubsec:n5_specialization}
\begin{figure}[t]
\centering
\includegraphics[width=1\linewidth]{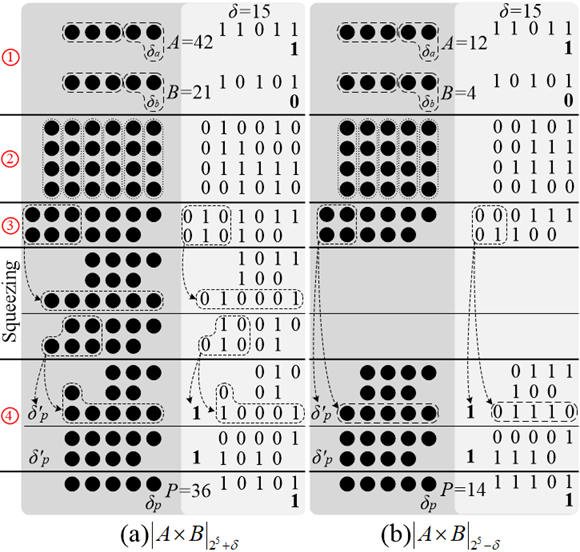}
\caption{Overall architecture of the proposed modulo-$(2^5 \pm \delta)$ multiplier. 
The example uses $\delta=15$ for illustration. 
(a) Modulo-$(2^5+\delta)$ with $A=42$ and $B=21$. 
(b) Modulo-$(2^5-\delta)$ with $A=12$ and $B=4$.}

\label{fig:overall_structure_for_n_5}
\vspace{-15pt}
\end{figure}

For \(n=5\), the number of operand groups in Stage~\redcircle{1} is given by
$
\Gamma = 1 + \left\lceil \frac{n-2}{3} \right\rceil
       = 1 + \left\lceil \frac{3}{3} \right\rceil
       = 2 .
$
Accordingly, each operand is decomposed into two groups: the least-significant group
\(g^{A}_{0}\), which includes the twit and the two least-significant bits, and one
3-bit group \(g^{A}_{1}\) containing the remaining bits \((a_4,a_3,a_2)\).
The same grouping is applied to operand \(B\).
The details of this operation are illustrated in Fig.~\ref{fig:overall_structure_for_n_5}(a) and Fig.~\ref{fig:overall_structure_for_n_5}(b) for moduli of the form \(2^n+\delta\) and
\(2^n-\delta\), respectively.
Based on this grouping, Stage~\redcircle{2} generates \(\Gamma^2 = 4\) modular partial
products per modulus channel.

Stage~\redcircle{3} performs multi-operand reduction by employing a 4:2 compressor to accumulate the generated partial products into two carry-save operands. In the present \(n=5\) case study, the reduced output width is \(n+1\) bits for moduli of the form \(2^n-\delta\), so the reduced operands can be processed directly by Stage~\redcircle{4} without any squeezing operation. 
However, for moduli of the form \(2^n+\delta\), the reduced output width becomes \(n+2\) bits. Therefore,
to make it compatible with Stage~\redcircle{4},
an additional squeezing operation is applied before the final modular addition.

Finally, Stage~\redcircle{4} performs the final modular addition. Although the overall computation flow is the same for both \(2^n-\delta\) and \(2^n+\delta\) moduli, minor implementation differences arise from their distinct representation formats. The computation then proceeds exactly as described in Section~\ref{subsec:generic_alg_arch} for each modulus channel \(m_i \in \mathcal{M}\).

\vspace{5pt}
\noindent
\textbf{Example 3:} In Fig.~\ref{fig:overall_structure_for_n_5}, the numerical examples demonstrate the
operation of the proposed architecture for \(n=5\) and \(\delta=15\). 

\begin{itemize}
    \item In Fig.~\ref{fig:overall_structure_for_n_5}(a), the modulus is \(m_i=2^5+15=47\), with \(A=42\) and \(B=21\), giving 
    \[ P = |42 \times 21|_{47} = 36.\]
    \item In Fig.~\ref{fig:overall_structure_for_n_5}(b), the modulus is \(m_i=2^5-15=17\), with \(A=12\) and \(B=4\), giving 
    \[ P = |12 \times 4|_{17} = 14.\]
\end{itemize}
\vspace{5pt}
\noindent
The detailed intermediate datapath values for both cases are shown in
Fig.~\ref{fig:overall_structure_for_n_5}.

The representative \(n=5\) case illustrates how the proposed framework is instantiated in a practical high-dynamic-range setting. Nevertheless, the framework is not limited to this specific channel width. To establish its broader applicability, Section~\ref{sec:eval} presents additional evaluations for larger channel widths, namely \(n=8\) and \(n=11\), using representative values of \(\delta\). That section also reports the corresponding circuit-level and system-level results.

\section{Performance Evaluation and Comparisons}
\label{sec:eval}

This section evaluates the proposed twit-based modulo-$(2^n \pm \delta)$ multiplier from complementary perspectives. We first describe the evaluation methodology and the comparison setup used throughout the study. We then present an analytical evaluation to reveal the main scaling trends and identify the dominant delay and hardware-cost components of the compared architectures. Next, we report circuit-level synthesis results obtained under a uniform CMOS flow, providing a more realistic assessment of delay, area, and power. Finally, we present a system-level case study to examine how the circuit-level benefits of the proposed approach translate into lower end-to-end latency over a broad range of modular multiplication and addition workloads.

\subsection{Methodology}
\label{subsec:eval_methodology}

The evaluation is carried out using two complementary approaches. First, analytical estimates are used to obtain a coarse understanding of architectural scaling, critical-path composition, and hardware-cost trends. Second, detailed synthesis results are reported for delay, area, and power under a uniform implementation flow.

For comparison, we implemented the proposed design in HDL together with the most relevant generic RNS-oriented baselines available in the literature, namely the designs of Hiasat~\cite{hiasat2000new} and Matutino \emph{et al.}~\cite{matutino2012rns}. All designs were verified using extensive random test vectors together with manually generated corner-case vectors to ensure functional correctness. In addition, post-synthesis verification was performed to confirm that the synthesized netlists preserve the intended behavior.

All HDL descriptions were synthesized using Synopsys Design Compiler with the FreePDK 45\,nm technology under identical settings to ensure a fair comparison. Although the proposed framework is defined generically for moduli of the form \(2^n \pm \delta\), exhaustive evaluation over all admissible values of \(\delta\) becomes increasingly expensive as \(n\) grows. Therefore, the detailed \(n=5\) case study is complemented by additional evaluations for larger channel widths, namely \(n=8\) and \(n=11\), using representative values of \(\delta\). This allows the evaluation to capture both the fine-grained behavior of the worked example and the broader applicability of the proposed approach.

\begin{table*}[t]
\centering
\caption{Block-level analytical composition of the compared generic modulo-$(2^n \pm \delta)$ multipliers.}
\label{tab:analytical_structure}
\renewcommand{\arraystretch}{1.2}
\setlength{\tabcolsep}{3pt}
\footnotesize

\begin{tabular}{
>{\raggedright\arraybackslash}p{0.12\textwidth}
>{\centering\arraybackslash}p{0.16\textwidth}
>{\centering\arraybackslash}p{0.16\textwidth}
>{\centering\arraybackslash}p{0.16\textwidth}
>{\centering\arraybackslash}p{0.16\textwidth}
>{\centering\arraybackslash}p{0.16\textwidth}
}
\toprule
\makecell{Building block}
& \makecell{Fig.~1(a)\\Modulo-$(2^n-\delta)$}
& \makecell{Fig.~1(b)\\Modulo-$(2^n-\delta)$}
& \makecell{Fig.~1(b)\\Modulo-$(2^n+\delta)$}
& \makecell{Proposed\\Modulo-$(2^n-\delta)$}
& \makecell{Proposed\\Modulo-$(2^n+\delta)$}
\\
\midrule

\makecell[l]{Mul Bin$(n\times n)$}
& $\mathbf{1}$
& $\mathbf{1}$
& $\mathbf{1}$
& $-$
& $-$
\\

\makecell[l]{CM$(i\times c)$}
& \makecell{$\bm{((n-1-p_H)\times p_H)}$}
& \makecell{$\bm{(n\times p_S)}+\bm{(p_S\times p_S)}$}
& \makecell{$\bm{(n\times p_S)}+\bm{(p_S\times p_S)}$}
& $-$
& $-$
\\

\makecell[l]{CL(input)}
& \makecell{$(p_H+2)$$+\mathbf{(2)}$}
& \makecell{$\mathbf{(4)}$$+\mathbf{(2)}$}
& \makecell{$(2)+\mathbf{(4)}$$+\mathbf{(2)}$}
& \makecell{$(\Gamma^2-1)(6)+\bm{\mathbf{(6)}}+$\\$\bm{(2\lambda+2)}+\mathbf{(2)}$}
& \makecell{$(\Gamma^2-1)(6)+\mathbf{(6)}+$\\$\bm{(2\lambda+4)}+\mathbf{(2)}$}
\\

CSA
& $\mathbf{1}$
& $\mathbf{2}$
& $\mathbf{3}$
& $\bm{\lambda+1}$
& $\bm{\lambda+1}$
\\

\makecell[l]{CPA$(n\text{-bit})$}
& $\mathbf{2}$
& $\mathbf{1}+1$
& $\mathbf{1}+1$
& $\mathbf{1}$
& $\mathbf{1}$
\\

\makecell[l]{MUX$(n\text{-bit})$}
& $-$
& \makecell{$4{:}1+\mathbf{4{:}1}+\mathbf{2{:}1}$}
& \makecell{$4{:}1+\mathbf{4{:}1}+\mathbf{2{:}1}$}
& $-$
& $-$
\\

\bottomrule
\end{tabular}
\end{table*}

\subsection{Analytical Evaluation}
\label{subsec:eval_analytical}

The analytical evaluation aims to expose the main structural differences between the proposed architecture and prior generic modulo multipliers without relying on a specific technology library. To this end, each design is decomposed into its fundamental building blocks, and the resulting block-level composition is summarized in Table~\ref{tab:analytical_structure}. The considered blocks include \(n\times n\) binary multipliers (Mul Bin), \(n\times c\) constant multipliers (CM), \(n\)-input combinational logic blocks (CL), Carry-Save Adders (CSA), Carry-Propagate Adders (CPA), and \(2{:}1\) and \(4{:}1\) multiplexers. Table~\ref{tab:analytical_structure} also identifies the blocks contributing to the critical-path delay, highlighted in boldface.

Delay and hardware cost are evaluated in terms of \(\Delta G\) and \(\#G\), representing the delay and hardware cost of a simple two-input logic gate, respectively. For delay analysis, XOR gates and multiplexers are modeled with a delay of \(2\Delta G\), while an \(n\)-bit CSA incurs \(4\Delta G\). The delay of an \(n\)-bit CPA implemented using a Kogge--Stone parallel-prefix structure is modeled as
$
(3+2\lceil \log_2 n\rceil)\Delta G.
$
For hardware-cost analysis, XOR gates and multiplexers require \(3\#G\), while CSA structures require \(9\#G\). The hardware cost of an \(n\)-bit CPA is modeled as
$
(3+3n\lceil \log_2 n\rceil-3n)\#G.
$
For an \(n\)-input combinational logic block, the delay is modeled as
$
\lceil\log_2 n\rceil\Delta G,
$
and the corresponding hardware cost is assumed to be \(n\#G\).

The binary multiplier is modeled as a three-stage structure consisting of partial-product generation, multi-operand reduction, and a final carry-propagate addition. In contrast, the constant multiplier does not require a partial-product generation stage, since one operand is fixed. All final additions are assumed to be implemented using Kogge--Stone parallel-prefix adders.

In the proposed architecture, the partial-product generation stage consists of \(\Gamma^2\) local 6-input logic blocks, where only one contributes to the critical path. The reduction stage is implemented using a \(\lambda\)-level CSA tree, followed by one additional CSA in the final modular addition stage. Furthermore, the modular correction stage requires a \((2\lambda+2)\)-input combinational logic block for moduli of the form \(2^n-\delta\), and a \((2\lambda+4)\)-input combinational logic block for moduli of the form \(2^n+\delta\), followed by a final XOR-based twit correction.

\begin{figure}[t]
\centering
\includegraphics[width=1.0
\linewidth]{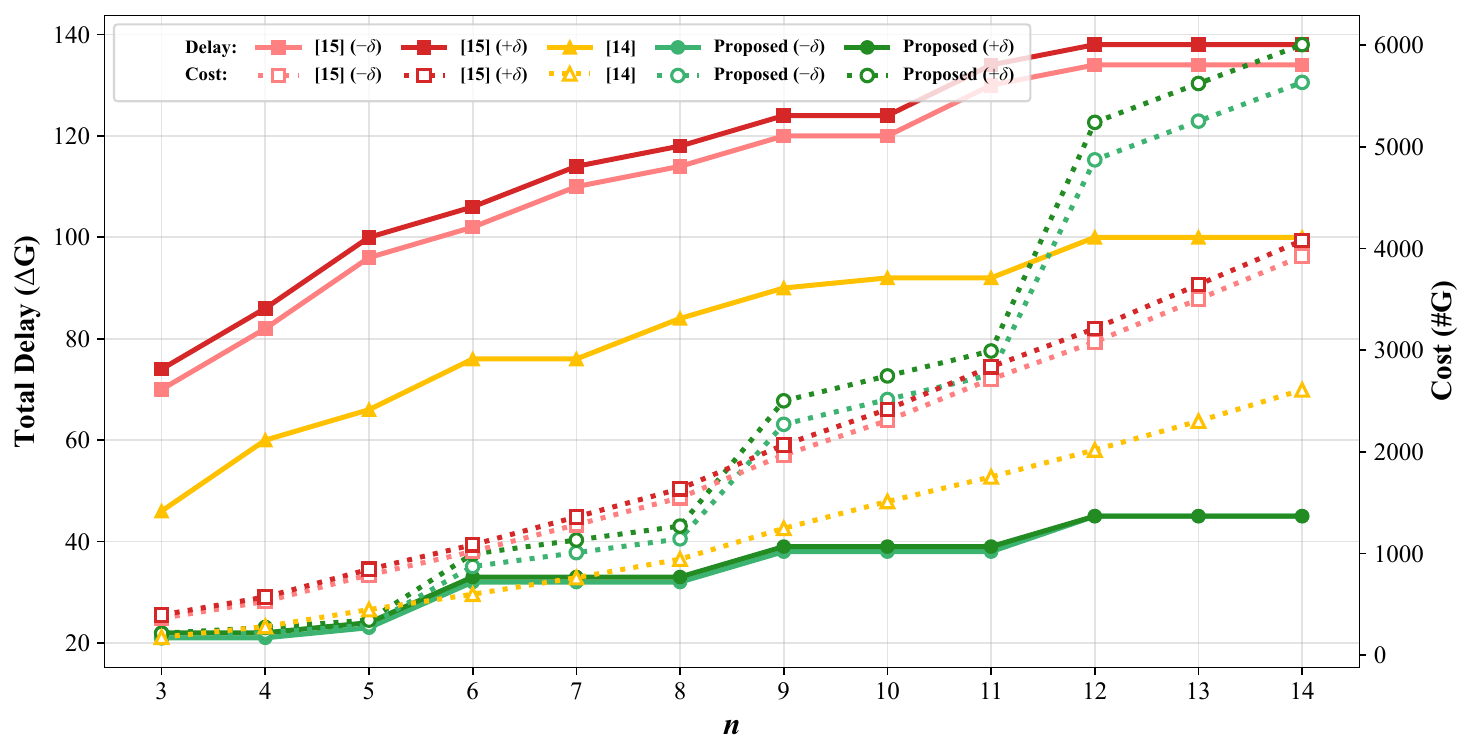}
\caption{Analytical delay ($\Delta G$) and hardware-cost ($\#G$) trends of the compared generic modulo-$(2^n \pm \delta)$ multipliers for ($3 \le n \le 16$).}
\label{fig:analytical_compare}
\end{figure}

Using the block-level decomposition of Table~\ref{tab:analytical_structure}, together with the corresponding delay and hardware-cost models, the channel-width parameter \(n\) is varied from 3 to 16. The resulting analytical delay and hardware-cost trends are illustrated in Fig.~\ref{fig:analytical_compare}.
As highlighted in Table~\ref{tab:analytical_structure}, unlike prior designs that rely on a full binary multiplier together with one or two constant multipliers and mux-intensive multi-input adder structures, the proposed architecture consists of only three explicit datapath components: modular partial-product generation using local 6-input logic, multi-operand reduction using a structured compressor tree, and a single carry-propagate adder in the final modular addition stage.

This structural difference is directly reflected in the analytical trends shown in Fig.~\ref{fig:analytical_compare}, where the proposed architecture demonstrates a more favorable scaling behavior in delay. Nevertheless, due to the quadratic growth of the number of partial products with respect to \(n\), the hardware cost of the proposed architecture increases more rapidly for larger channel widths. It should also be emphasized that the presented analytical evaluation provides only a first-order estimation of architectural behavior. Therefore, a more accurate comparison is performed through post-synthesis evaluation under a uniform implementation flow, as discussed in Section~\ref{subsec:eval_synthesis}.

\subsection{Circuit-Level Synthesis Results}
\label{subsec:eval_synthesis}

While the analytical evaluation is useful for exposing structural trends, it does not capture technology-dependent effects such as logic mapping, fan-out, interconnect overhead, and gate sizing. To obtain a more realistic comparison, all architectures were synthesized under a uniform CMOS implementation flow using Synopsys Design Compiler with the FreePDK 45\,nm standard-cell library. Unless otherwise stated, all reported results correspond to the typical process corner, nominal supply voltage, and a temperature of 25\,$^\circ$C.

Since publicly available RTL implementations of the prior generic modulo multipliers were not available, the baseline architectures in~\cite{hiasat2000new,matutino2012rns} were re-implemented in Verilog strictly according to their published formulations. Each implementation was carefully optimized, functionally verified using extensive random and corner-case test vectors, and validated prior to synthesis.

The synthesis study first investigates the representative \(n=5\) case in detail and then extends the evaluation to larger channel widths, namely \(n=8\) and \(n=11\), using representative values of \(\delta\). For each configuration, delay, area, and power are evaluated under identical synthesis settings. These experiments verify whether the trends predicted by the analytical model remain valid after technology mapping and quantify the actual circuit-level benefits of the proposed architecture.

\begin{figure}[t]
\centering

\subfloat[Delay (minimum achievable) (ns)\label{fig:delay}]{
    \includegraphics[width=0.95\columnwidth,height=0.17\textheight]{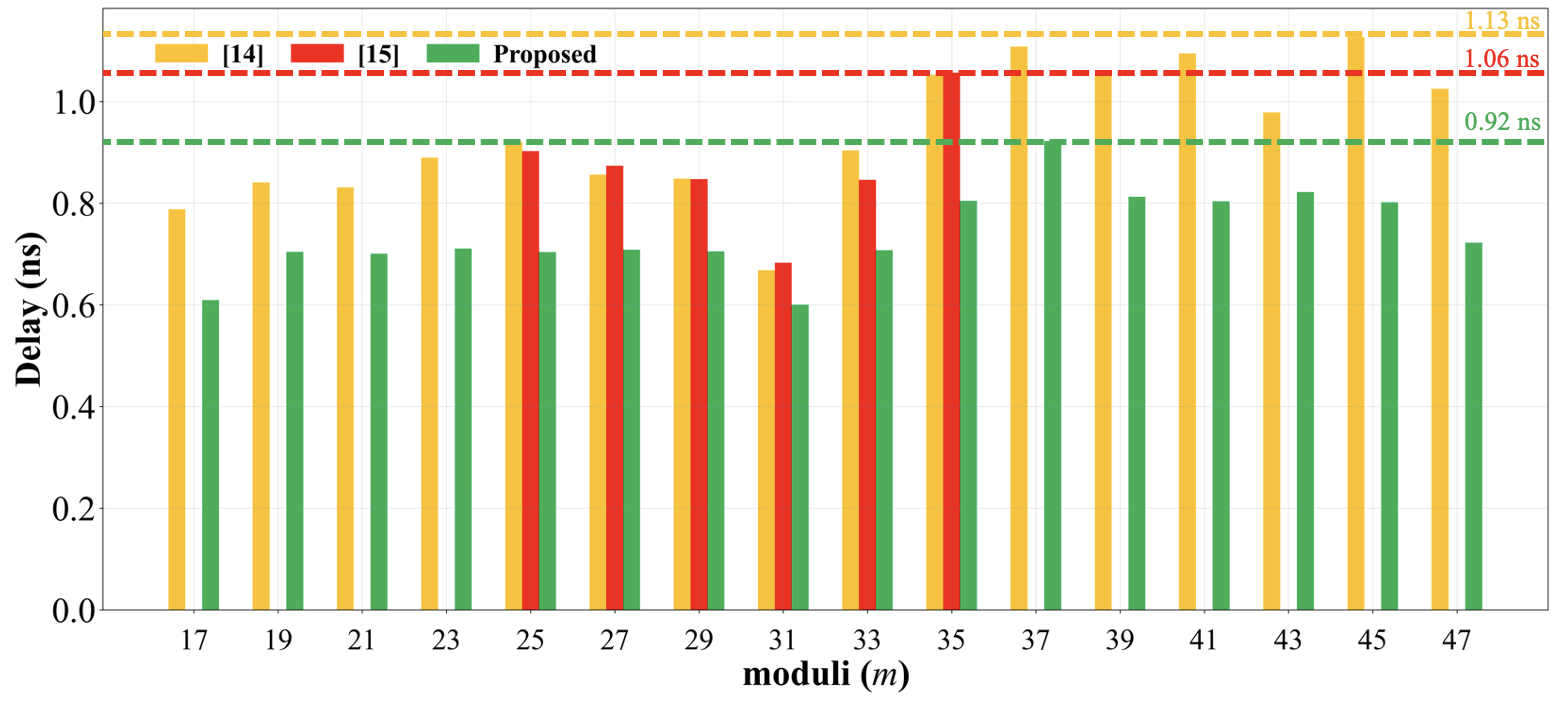}
}

\vspace{-2pt}

\subfloat[Area ($\mu m^2$)\label{fig:area}]{
    \includegraphics[width=0.95\columnwidth,height=0.17\textheight]{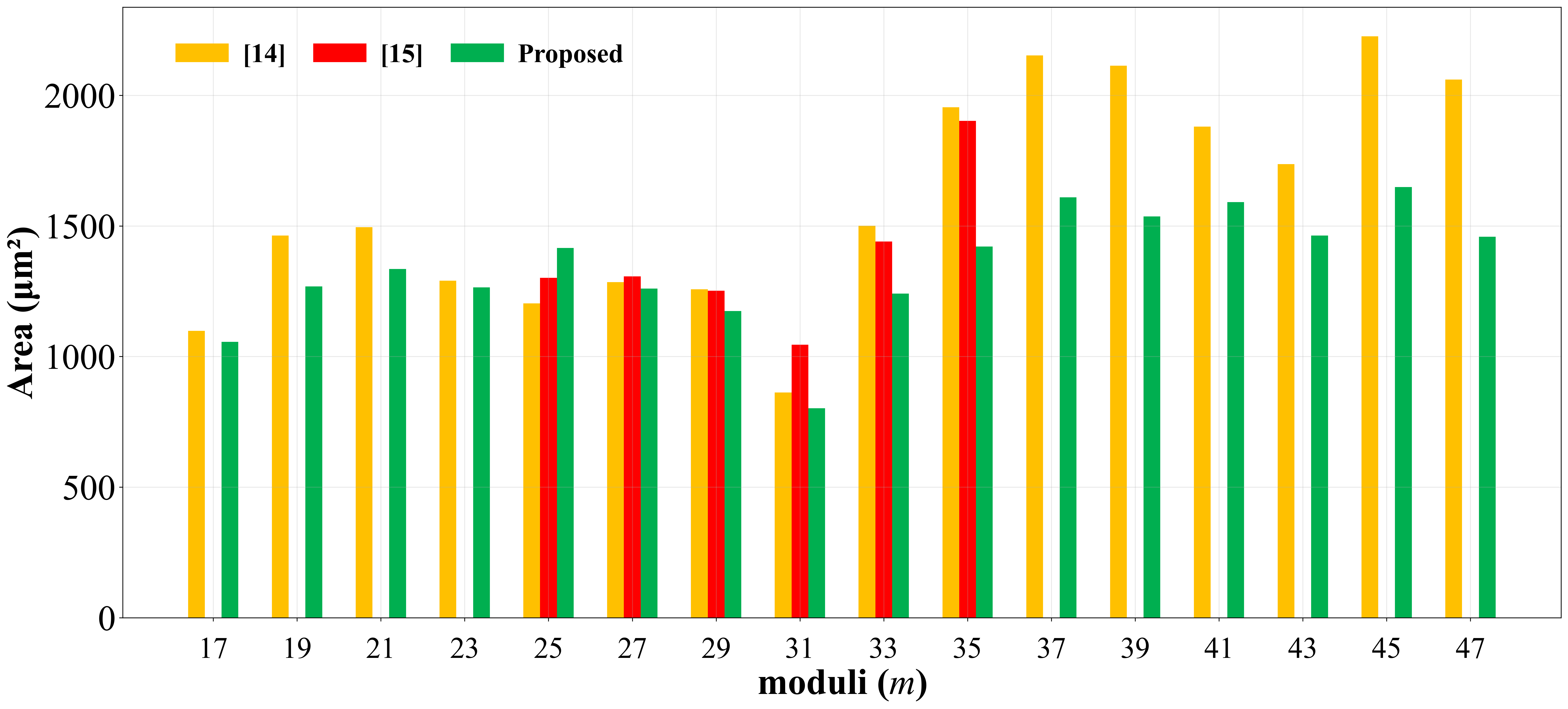}
}

\vspace{-2pt}

\subfloat[Power (mW)\label{fig:power}]{
    \includegraphics[width=0.95\columnwidth,height=0.17\textheight]{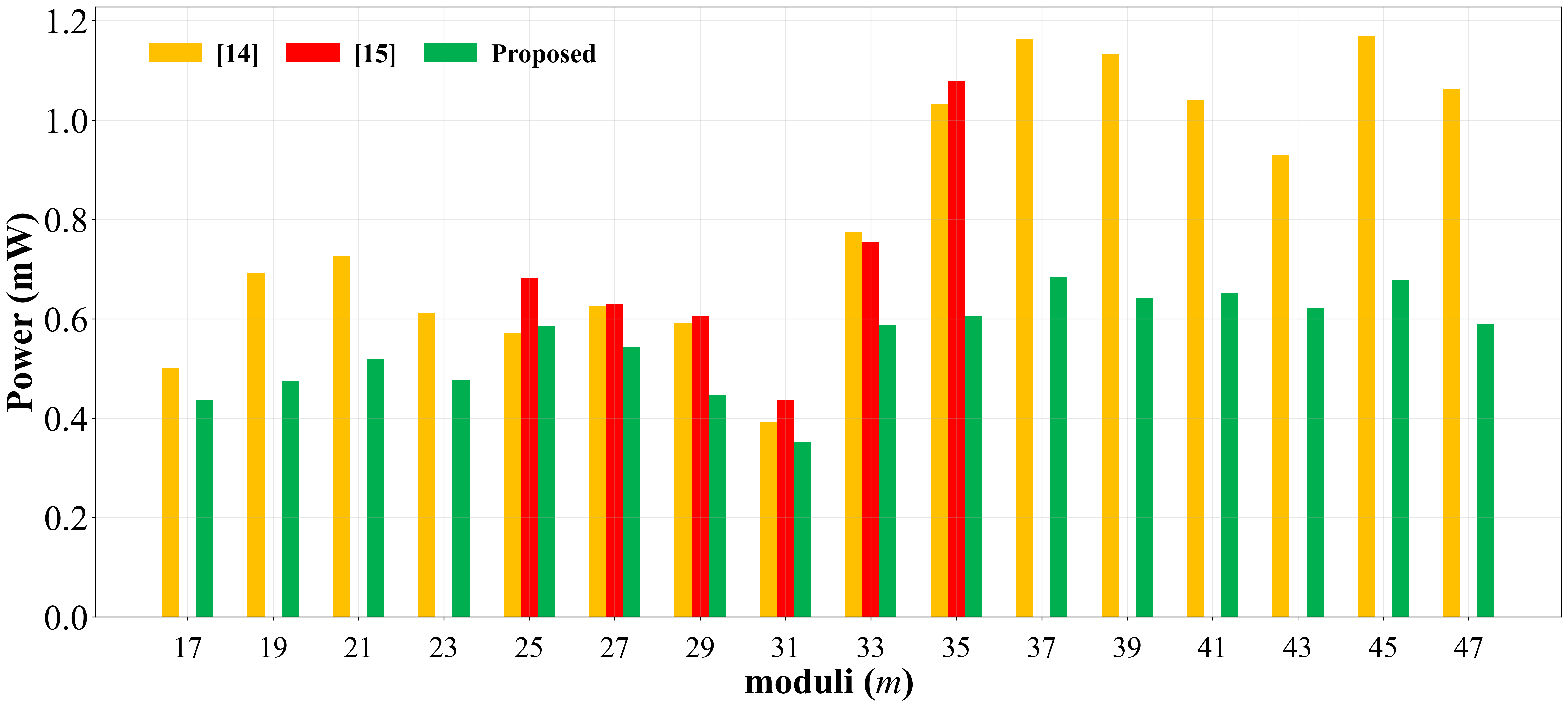}
}

\caption{Synthesis comparison for the representative \(n=5\) moduli.}
\label{fig:synthesis_n5}
\vspace{-10pt}
\end{figure}

Timing closure was performed independently for each architecture in order to determine its minimum achievable critical-path delay. For a fair comparison of area and power, however, all designs were additionally synthesized under an identical timing constraint corresponding to a common operating point. This ensures that hardware and energy metrics are evaluated at the same target performance level. The resulting delay, area, and power values for different moduli are summarized in Fig.~\ref{fig:synthesis_n5}.

Fig.~\ref{fig:synthesis_n5}(a) reports the minimum achievable delay for each architecture. Consistent with the analytical evaluation, the proposed design achieves shorter critical paths due to the elimination of intermediate carry-propagation stages and the dominance of carry-save computation throughout the datapath. The delay advantage becomes more pronounced for larger moduli, confirming the scalability of the proposed organization. On average, the proposed architecture achieves a 20.5\% delay reduction compared with the generic baseline of~\cite{hiasat2000new}.

Figs.~\ref{fig:synthesis_n5}(b) and~\ref{fig:synthesis_n5}(c) present the corresponding area and power results under a common timing constraint. Despite the additional local combinational logic introduced for modular partial-product generation, the proposed architecture achieves average improvements of 13.2\% in area and 28.0\% in power compared with~\cite{hiasat2000new}. These results indicate that the reduction in carry-propagation overhead effectively compensates for the added local logic complexity.

It is worth noting that the architecture in~\cite{matutino2012rns} exhibits structural constraints that restrict its applicability to only a subset of the evaluated moduli. Consequently, synthesis results for several moduli are unavailable for this design, which explains the missing entries (red bars) in Fig.~\ref{fig:synthesis_n5}. This limitation originates from the specific residue representation and reduction mechanism employed in~\cite{matutino2012rns}, which cannot be directly generalized to arbitrary moduli of the form \(2^n \pm \delta\).

\begin{figure*}[t]
\centering

\subfloat[\(m_i=25\)]{%
\includegraphics[width=0.32\textwidth]{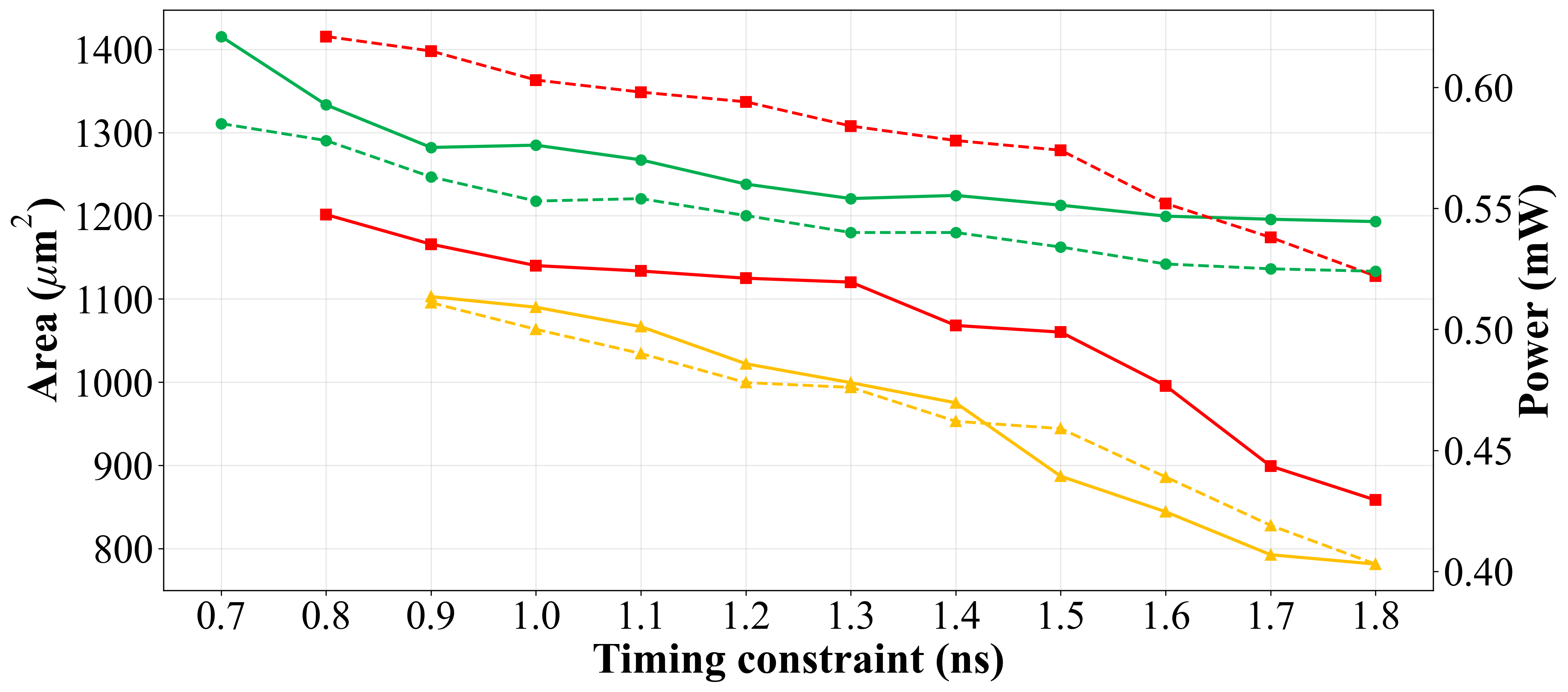}
\label{fig:eval_a}}
\hfill
\subfloat[\(m_i=27\)]{%
\includegraphics[width=0.32\textwidth]{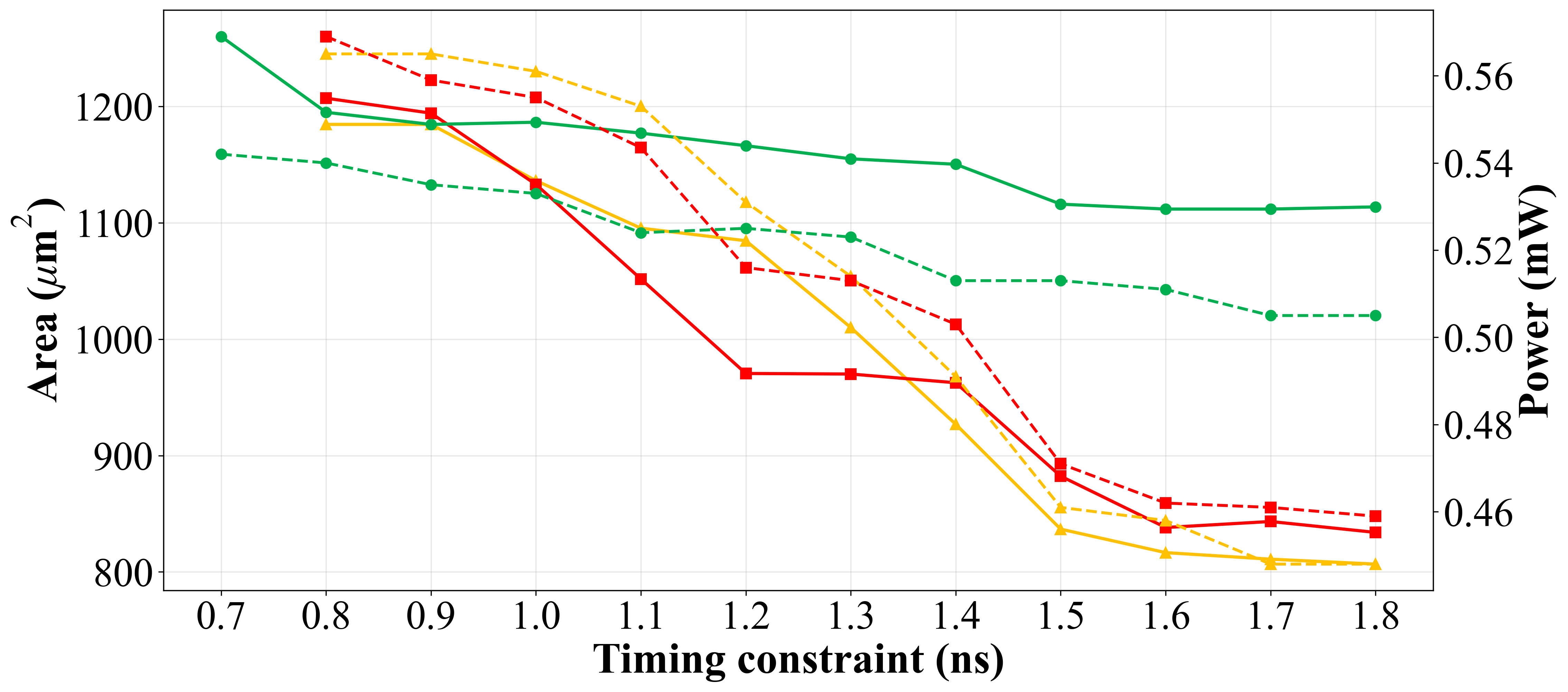}
\label{fig:eval_b}}
\hfill
\subfloat[\(m_i=29\)]{%
\includegraphics[width=0.32\textwidth]{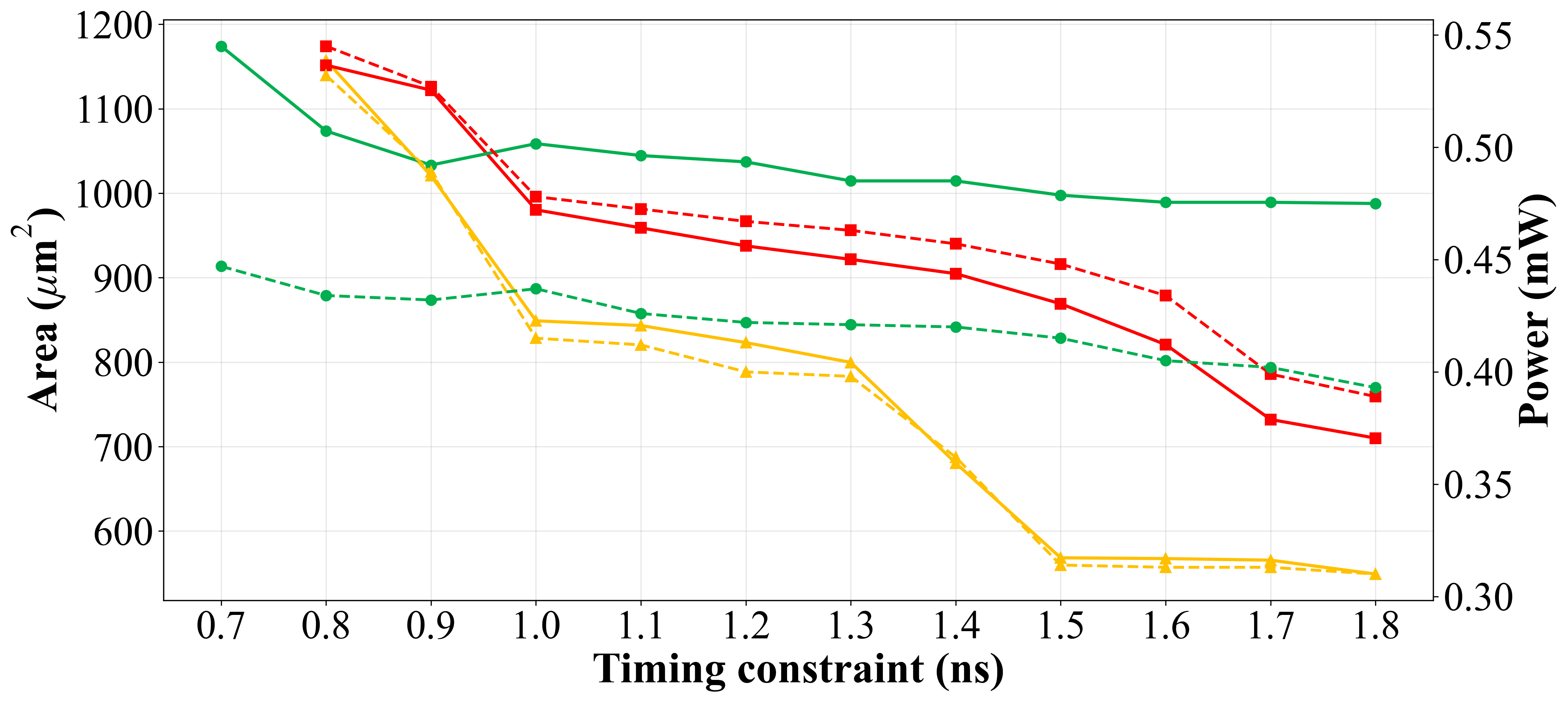}
\label{fig:eval_c}}

\vspace{4pt}

\subfloat[\(m_i=31\)]{%
\includegraphics[width=0.32\textwidth]{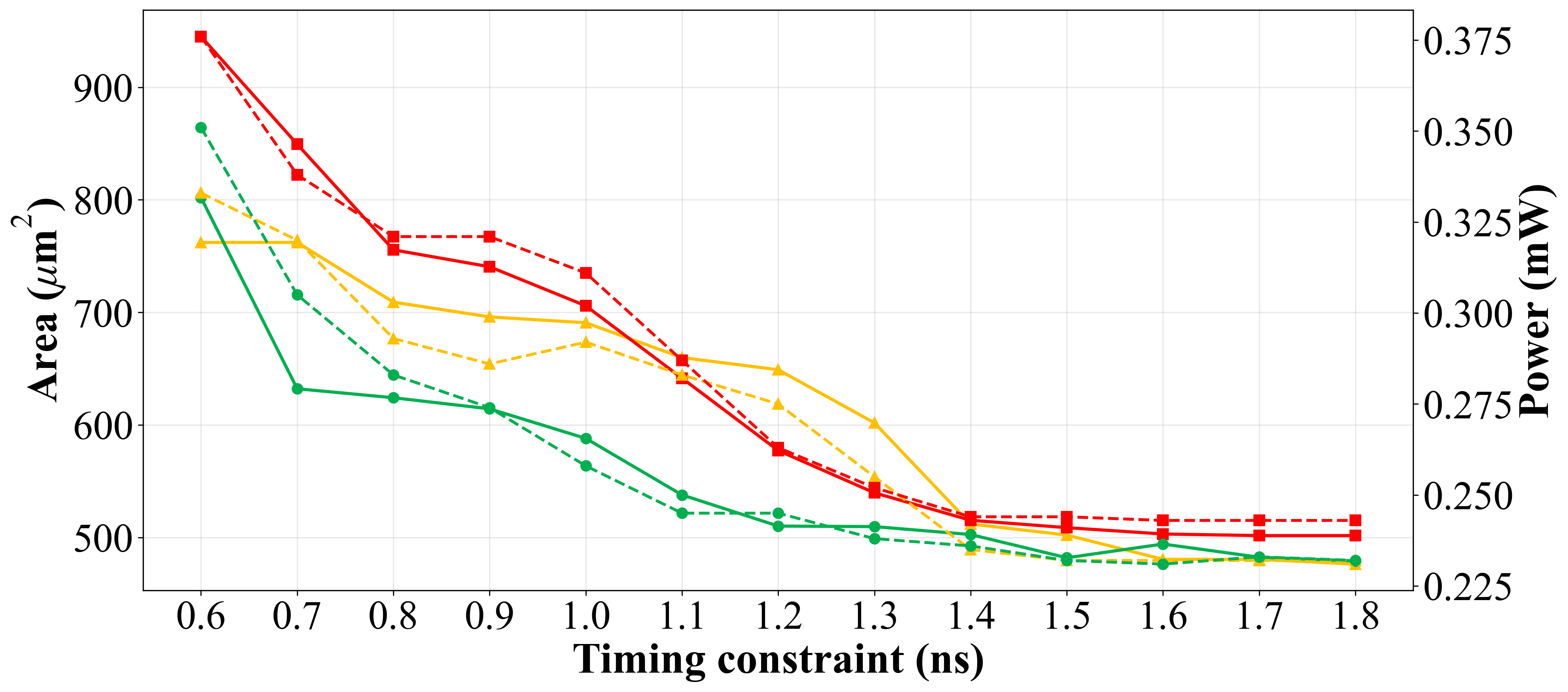}
\label{fig:eval_d}}
\hfill
\subfloat[\(m_i=33\)]{%
\includegraphics[width=0.32\textwidth]{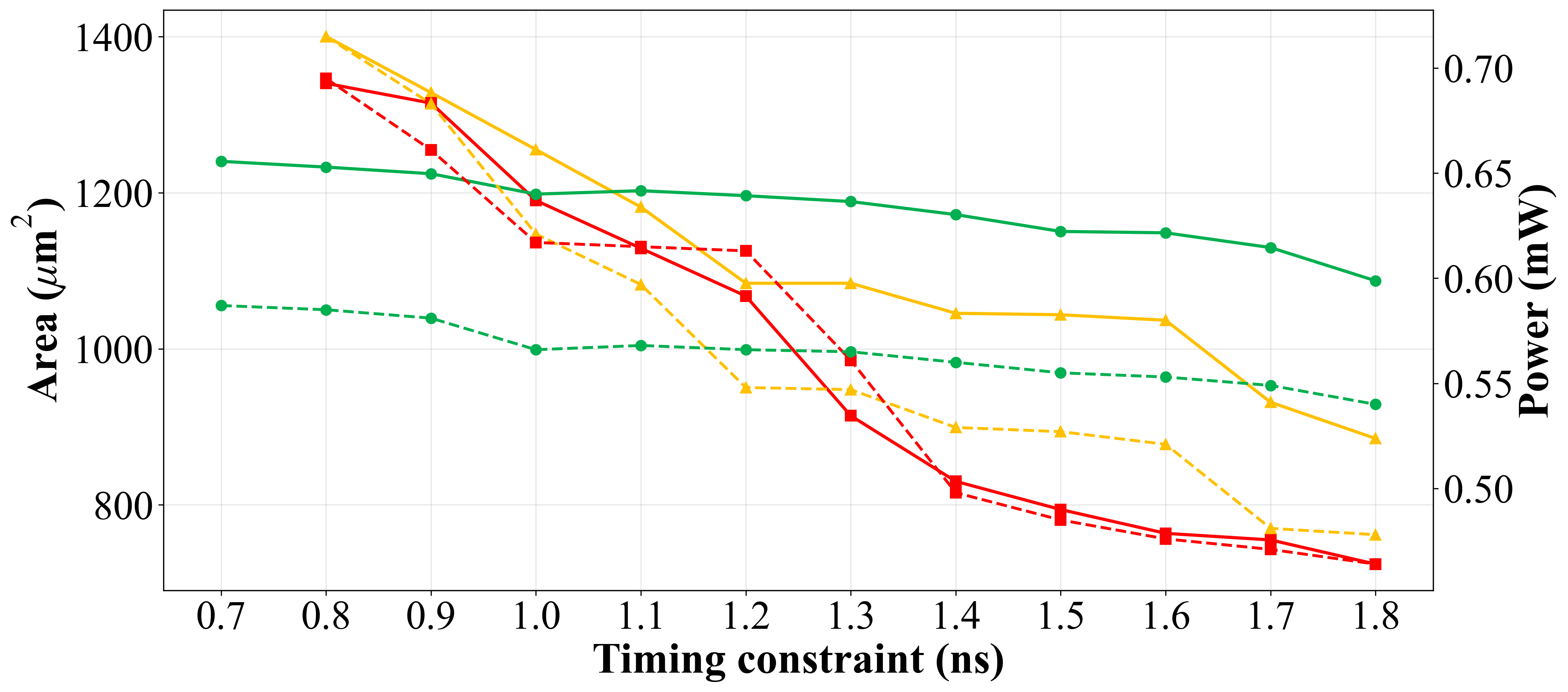}
\label{fig:eval_e}}
\hfill
\subfloat[\(m_i=35\)]{%
\includegraphics[width=0.32\textwidth]{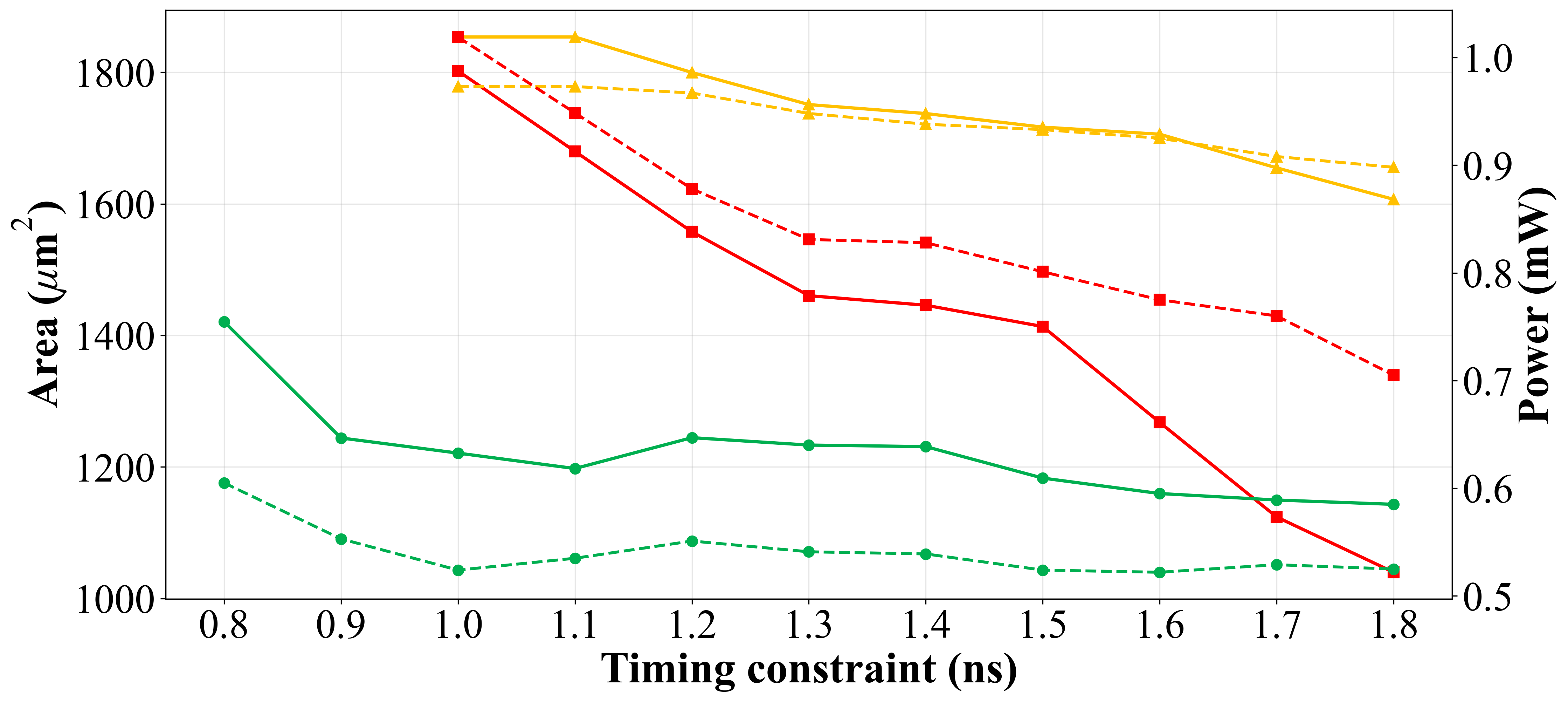}
\label{fig:eval_f}}

\includegraphics[width=0.65\textwidth]{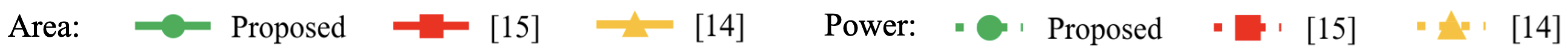}

\caption{Area and power of the moduli supported by all compared architectures under various timing constraints.}
\label{fig:synthesis_sweep}
\vspace{-8pt}
\end{figure*}

To further investigate the timing--area--power tradeoff, a more detailed evaluation was conducted for the six moduli \(25\), \(27\), \(29\), \(31\), \(33\), and \(35\), for which the architecture in~\cite{matutino2012rns} is also applicable. In this experiment, the synthesis timing constraint was swept 1.8\,ns down to the minimum achievable delay of each architecture. The resulting area and power trends are presented in Fig.~\ref{fig:synthesis_sweep}. The results again demonstrate the superiority of the proposed architecture, which consistently achieves lower area and power while also meeting more aggressive timing constraints than the competing designs. 

In practical residue-number-system implementations, the overall operating frequency is determined by the slowest modular channel, since faster channels do not provide any throughput advantage once the worst-case modulus defines the global critical path. Therefore, for each architecture, the maximum delay observed among all moduli in Fig.~\ref{fig:synthesis_n5}(a), indicated by the dashed line, was used as the synthesis timing constraint, and the synthesis process was repeated under this fixed delay target.

\begin{figure}[t]
\centering

\subfloat[Area $\times$ Delay ($\mu m^2$$\cdot$ns)\label{fig:area}]{
    \includegraphics[width=0.95\columnwidth,height=0.17\textheight]{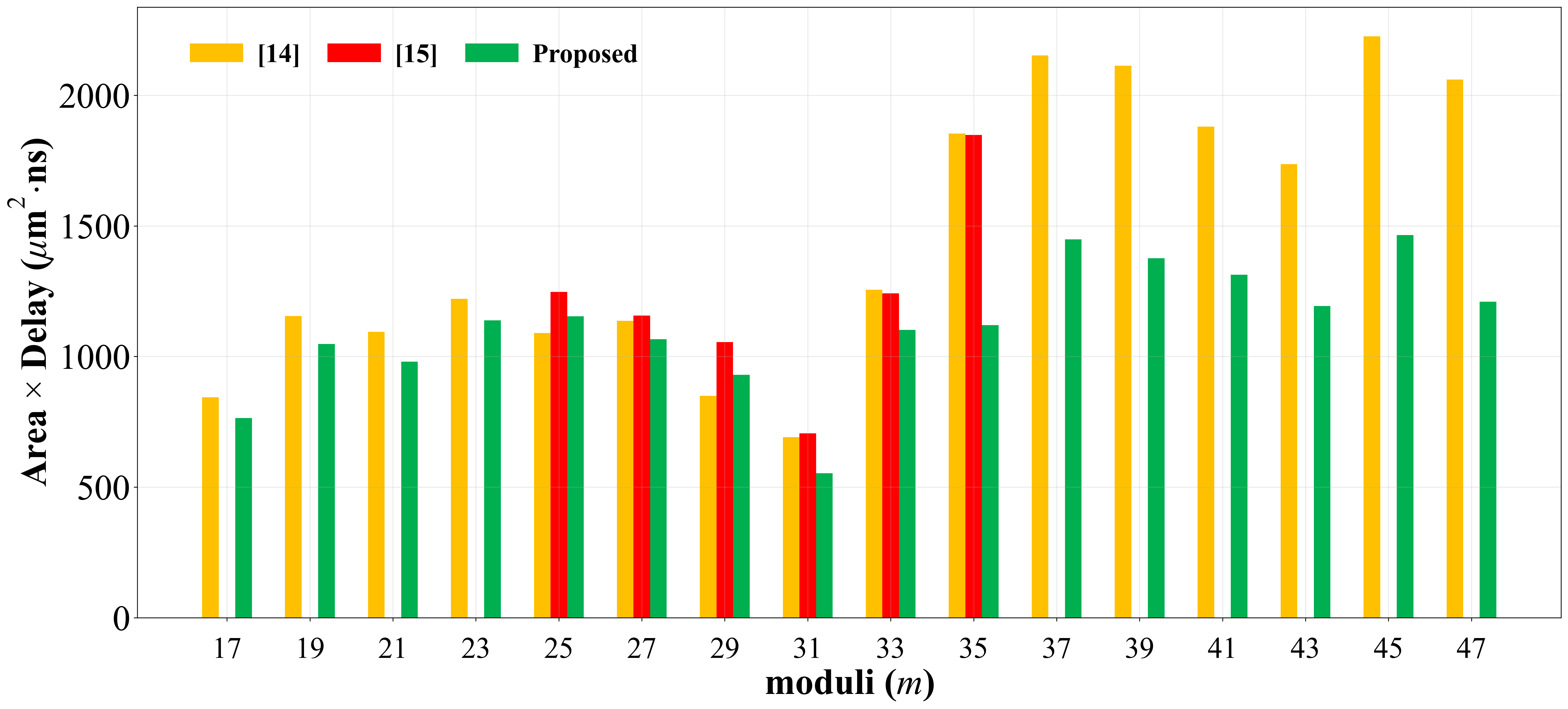}
}

\vspace{-2pt}

\subfloat[Power $\times$ Delay (mW$\cdot$ns)\label{fig:power}]{
    \includegraphics[width=0.92\columnwidth,height=0.17\textheight]{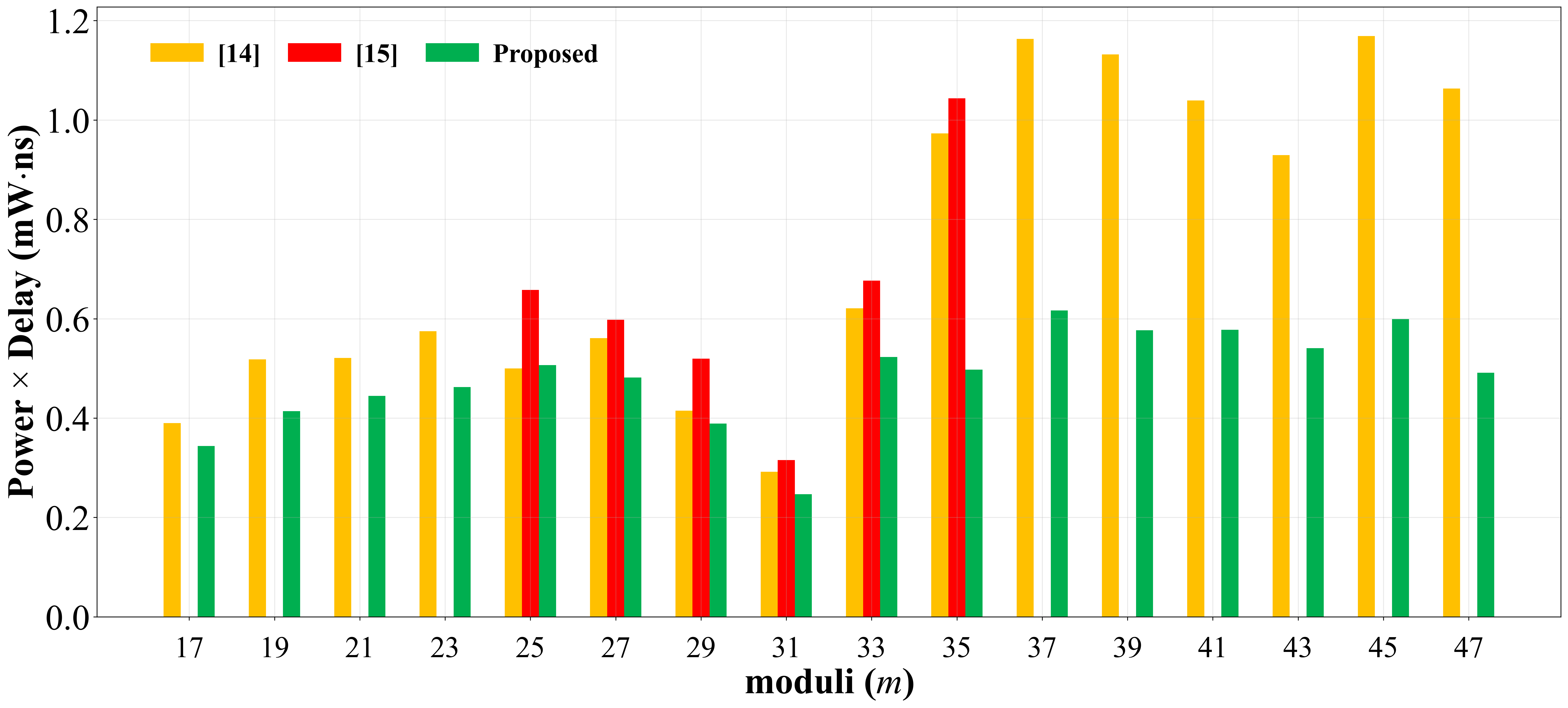}
}

\caption{Comparison of ADP and PDP at the timing constraint met by the slowest modulus.}
\label{fig:synthesis_worsecase}
\end{figure}

The resulting area-delay product (ADP) and power-delay product (PDP) values under this worst-case timing constraint are shown in Fig.~\ref{fig:synthesis_worsecase}. Even under a common worst-case delay target, the proposed architecture maintains superior efficiency in both area and power while simultaneously providing lower achievable delay. On average, the proposed design achieves approximately 19\% and 28\% improvements in area-delay and power-delay products, respectively.

The key results of this evaluation are further summarized in Table~\ref{tab:synth_best_points}. In this table, the proposed RNS configuration is also compared with the well-known three-modulus set of the form \(\tau=\{2^n,2^n\pm1\}\), with \(n=22\), as well as with a conventional binary implementation. The dynamic range is kept approximately the same across all cases to ensure a fair comparison. The results reported in Table~\ref{tab:synth_best_points} indicate the gains of our proposed design over the three-modulus and binary baselines. These results are subsequently used in the system-level evaluation presented in Section~\ref{subsec:eval_system}.

The synthesis results for larger channel widths, namely \(n=8\) and \(n=11\), are summarized in Table~\ref{tab:syn_8_11}. Since the number of admissible offsets grows significantly with \(n\), only representative values of \(\delta\) are selected for these larger-width cases. It should be noted that, for \(\delta=2^{n-1}-1\), either in the positive or negative form, the architecture in~\cite{matutino2012rns} is not applicable; therefore, the corresponding entries are omitted from the table.
Moreover, the reported ratios are normalized to the proposed design; therefore,
values larger than one indicate overhead relative to the proposed multiplier.

Consistent with the analytical model, \cite{hiasat2000new} can achieve lower
area or power in some \(2^n-\delta\) synthesis scenarios. However, this
advantage does not carry over to the complementary \(2^n+\delta\) cases, where
the same approach suffers from a substantial increase in implementation cost
because the effective datapath width must be increased. As a result, the
proposed architecture provides a more balanced implementation across both
\(2^n-\delta\) and \(2^n+\delta\) moduli.

Moreover, although the analytical model suggests a conservative area trend for
the proposed design, the synthesized results show that its area does not
increase significantly. This difference is mainly due to technology-dependent
optimization effects that are not captured by the analytical model, including
logic sharing, Boolean simplification, gate sizing, and technology mapping
performed by the synthesis tool.

\begin{table}[t]
\centering
\caption{Delay, area, and power comparison of design configurations with comparable dynamic-range coverage.}
\label{tab:synth_best_points}
\resizebox{\columnwidth}{!}{
\begin{tabular}{lccc}
\hline
\textbf{Design} & \textbf{Delay (ns)} & \textbf{Area ($\mu m^2$)} & \textbf{Power ($\mu$W)} \\
\hline
Proposed & 0.92 & 1609.70 & 685 \\
~\cite{hiasat2000new} & 1.13 & 2225.93 & 1169 \\
3-modulus set \(\tau\) & 2.10 & 15974.64 & 13331 \\
Conv. Binary & 3.22 & 32043.63 & 31593 \\
\hline
\end{tabular}}
\vspace{-6pt}
\end{table}

\subsection{Application-Level Evaluation}
\label{subsec:eval_system}

To provide a comprehensive system-level evaluation that is not tied to a single application kernel, we developed an in-house evaluation tool that estimates the execution delay of an arithmetic datapath based on  the synthesis delay of each modular arithmetic unit, the number of modular multiplications and additions. The tool also accounts for the delay overhead of the forward and reverse conversion units, so that the end-to-end impact of using the proposed RNS-based datapath can be assessed more realistically.

\begin{figure}[t]
\centering
\includegraphics[width=1.00
\linewidth]{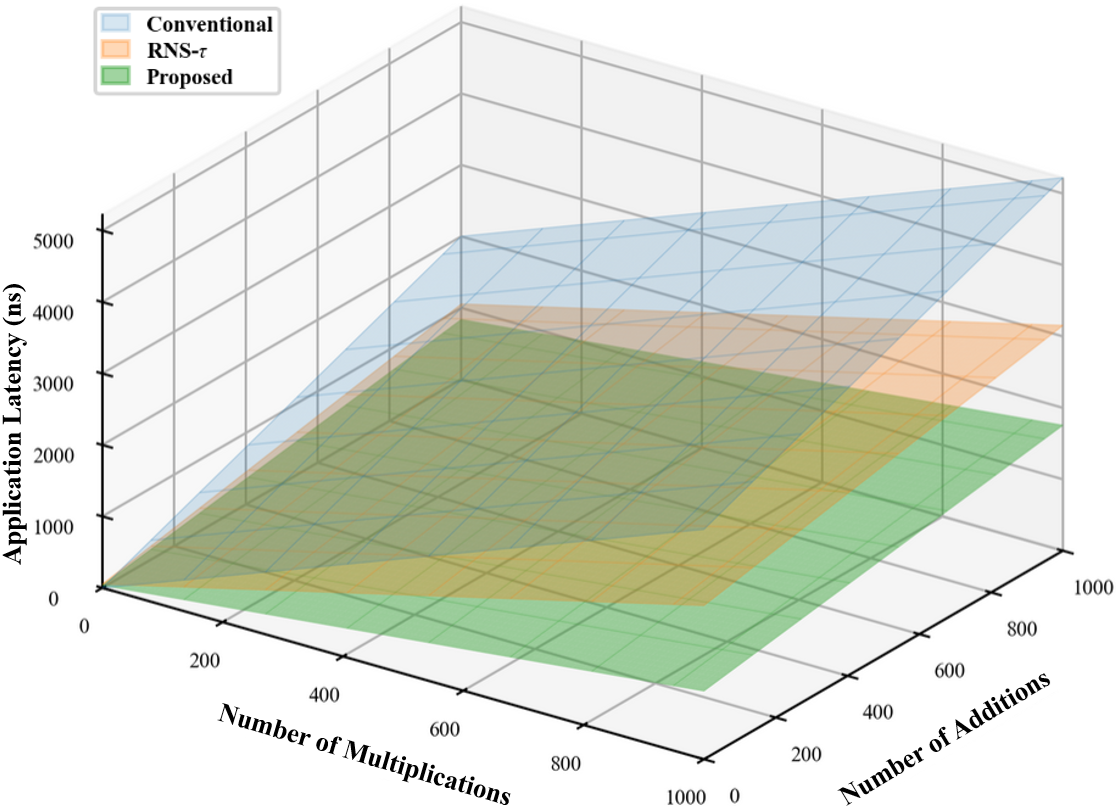}
\caption{Application-level delay evaluation as a function of the number of modular multiplications and additions.}
\label{fig:3d}
\end{figure}

The result of this evaluation is shown in Fig.~\ref{fig:3d} as a three-dimensional delay surface. The two horizontal axes represent the number of multiplication and addition operations, while the vertical axis shows the total execution delay. As can be observed, the surface corresponding to the proposed architecture remains below those of the conventional binary implementation and the well-known three-modulus RNS set. This improvement is mainly due to the fact that the proposed framework operates over smaller-width modular channels while still providing a comparable dynamic range. Consequently, the proposed design achieves the lowest overall delay across a wide range of multiplication- and addition-dominated workloads.

\begin{table*}[t]
\centering
\caption{Synthesis results of the compared generic modulo-\((2^n \pm \delta)\) multipliers for \(n=8\) and \(n=11\).}
\label{tab:syn_8_11}
\begin{tabular}{cccccccccccc}
\hline
$n$ & $\delta$ & $M$ & Design & Delay (ns) & Ratio & Area ($\mu$m$^2$) & Ratio & Power ($\mu$W) & Ratio & PDP ($\mu$W$\cdot$ns) & Ratio \\
\hline
\multirow{16}{*}{8} & \multirow{3}{*}{$-3$} & \multirow{3}{*}{253} & \cite{hiasat2000new} & 1.19 & 1.07 & \textbf{2010.48} & \textbf{0.75} & \textbf{1272} & \textbf{0.75} & \textbf{1513.55} & \textbf{0.81} \\
& & & \cite{matutino2012rns} & 1.32 & 1.19 & 2964.10 & 1.11 & 2221 & 1.31 & 2938.61 & 1.57 \\
& & & Proposed & \textbf{1.11} & \textbf{1.00} & 2679.70 & 1.00 & 1689 & 1.00 & 1871.75 & 1.00 \\
\cline{2-12}
& \multirow{3}{*}{$+3$} & \multirow{3}{*}{259} & \cite{hiasat2000new} & 1.70 & 1.40 & 9140.09 & 3.11 & 3374 & 1.61 & 5731.41 & 2.26 \\
& & & \cite{matutino2012rns} & 1.36 & 1.12 & 3435.28 & 1.17 & 2685 & 1.28 & 3646.77 & 1.44 \\
& & & Proposed & \textbf{1.21} & \textbf{1.00} & \textbf{2938.76} & \textbf{1.00} & \textbf{2094} & \textbf{1.00} & \textbf{2534.16} & \textbf{1.00} \\
\cline{2-12}
& \multirow{3}{*}{$-9$} & \multirow{3}{*}{247} & \cite{hiasat2000new} & 1.31 & 1.16 & 2592.41 & 1.01 & \textbf{1668} & \textbf{0.98} & 2188.75 & 1.13 \\
& & & \cite{matutino2012rns} & 1.26 & 1.12 & 2774.03 & 1.08 & 1994 & 1.17 & 2515.63 & 1.30 \\
& & & Proposed & \textbf{1.13} & \textbf{1.00} & \textbf{2558.15} & \textbf{1.00} & 1709 & 1.00 & \textbf{1931.68} & \textbf{1.00} \\
\cline{2-12}
& \multirow{3}{*}{$+9$} & \multirow{3}{*}{265} & \cite{hiasat2000new} & 1.59 & 1.22 & 8061.17 & 2.04 & 3198 & 1.15 & 5072.03 & 1.41 \\
& & & \cite{matutino2012rns} & 1.49 & 1.14 & \textbf{3882.52} & \textbf{0.98} & 3052 & 1.10 & 4549.31 & 1.26 \\
& & & Proposed & \textbf{1.30} & \textbf{1.00} & 3953.38 & 1.00 & \textbf{2769} & \textbf{1.00} & \textbf{3608.84} & \textbf{1.00} \\
\cline{2-12}
& \multirow{2}{*}{$-127$} & \multirow{2}{*}{129} & \cite{hiasat2000new} & 1.20 & 1.19 & \textbf{2351.66} & \textbf{0.90} & 1574 & 1.00 & 1886.60 & 1.20 \\
& & & Proposed & \textbf{1.01} & \textbf{1.00} & 2622.45 & 1.00 & \textbf{1569} & \textbf{1.00} & \textbf{1577.79} & \textbf{1.00} \\
\cline{2-12}
& \multirow{2}{*}{$+127$} & \multirow{2}{*}{383} & \cite{hiasat2000new} & 1.70 & 1.12 & 6750.88 & 1.52 & 2911 & 1.12 & 4951.03 & 1.26 \\
& & & Proposed & \textbf{1.52} & \textbf{1.00} & \textbf{4452.78} & \textbf{1.00} & \textbf{2594} & \textbf{1.00} & \textbf{3930.43} & \textbf{1.00} \\
\hline
\multirow{16}{*}{11} & \multirow{3}{*}{$-3$} & \multirow{3}{*}{2045} & \cite{hiasat2000new} & 1.60 & 1.20 & \textbf{3329.68} & \textbf{0.77} & \textbf{2230} & \textbf{0.73} & \textbf{3567.33} & \textbf{0.87} \\
& & & \cite{matutino2012rns} & 1.70 & 1.27 & 5136.02 & 1.18 & 4036 & 1.32 & 6851.11 & 1.68 \\
& & & Proposed & \textbf{1.34} & \textbf{1.00} & 4339.15 & 1.00 & 3050 & 1.00 & 4078.15 & 1.00 \\
\cline{2-12}
& \multirow{3}{*}{$+3$} & \multirow{3}{*}{2051} & \cite{hiasat2000new} & 2.21 & 1.57 & 54554.70 & 10.04 & 9283 & 2.18 & 20541.41 & 3.43 \\
& & & \cite{matutino2012rns} & 1.76 & 1.25 & 5989.68 & 1.10 & 5092 & 1.20 & 8969.05 & 1.50 \\
& & & Proposed & \textbf{1.41} & \textbf{1.00} & \textbf{5431.68} & \textbf{1.00} & \textbf{4250} & \textbf{1.00} & \textbf{5984.42} & \textbf{1.00} \\
\cline{2-12}
& \multirow{3}{*}{$-9$} & \multirow{3}{*}{2039} & \cite{hiasat2000new} & 1.70 & 1.21 & 4291.75 & 1.02 & \textbf{2836} & \textbf{0.95} & 4821.20 & 1.15 \\
& & & \cite{matutino2012rns} & 1.72 & 1.22 & 4676.57 & 1.12 & 3733 & 1.25 & 6418.52 & 1.53 \\
& & & Proposed & \textbf{1.41} & \textbf{1.00} & \textbf{4192.26} & \textbf{1.00} & 2985 & 1.00 & \textbf{4199.60} & \textbf{1.00} \\
\cline{2-12}
& \multirow{3}{*}{$+9$} & \multirow{3}{*}{2057} & \cite{hiasat2000new} & 2.35 & 1.56 & 64344.38 & 12.98 & 9338 & 2.47 & 21957.43 & 3.86 \\
& & & \cite{matutino2012rns} & 1.88 & 1.25 & 6302.23 & 1.27 & 5305 & 1.41 & 9965.44 & 1.75 \\
& & & Proposed & \textbf{1.51} & \textbf{1.00} & \textbf{4955.34} & \textbf{1.00} & \textbf{3775} & \textbf{1.00} & \textbf{5689.68} & \textbf{1.00} \\
\cline{2-12}
& \multirow{2}{*}{$-1023$} & \multirow{2}{*}{1025} & \cite{hiasat2000new} & 2.08 & 1.19 & 13422.91 & 2.01 & 5360 & 1.33 & 11157.90 & 1.58 \\
& & & Proposed & \textbf{1.75} & \textbf{1.00} & \textbf{6687.65} & \textbf{1.00} & \textbf{4042} & \textbf{1.00} & \textbf{7081.58} & \textbf{1.00} \\
\cline{2-12}
& \multirow{2}{*}{$+1023$} & \multirow{2}{*}{3071} & \cite{hiasat2000new} & 2.31 & 1.23 & 50857.13 & 7.14 & 7920 & 1.84 & 18273.02 & 2.27 \\
& & & Proposed & \textbf{1.87} & \textbf{1.00} & \textbf{7126.43} & \textbf{1.00} & \textbf{4300} & \textbf{1.00} & \textbf{8044.44} & \textbf{1.00} \\
\hline
\end{tabular}
\end{table*}


\section{Conclusion}
\label{sec:conclusion}

This paper introduced a twit-compatible generic hardware framework for modulo-$(2^n\pm\delta)$ multiplication.
By generating modular partial products and performing reduction through a structured compressor/squeezing flow, the proposed design provides an efficient and scalable alternative to prior arithmetic-based generic multipliers.
Circuit-level synthesis results show that the proposed multiplier achieves an average delay reduction of 20\%, together with 13\% and 28\% improvements in area and power, respectively, compared with a representative state-of-the-art generic design.
Application-level evaluation further shows that these circuit-level improvements translate into lower end-to-end delay across a broad range of multiplication- and addition-dominated workloads.
These results demonstrate that the proposed framework can support high-dynamic-range RNS computation using narrower residue channels and reduced modular-arithmetic latency.

\bibliographystyle{IEEEtran}
\bibliography{references}


\begin{IEEEbiography}[{\includegraphics[width=1in,height=1.25in,clip,keepaspectratio]{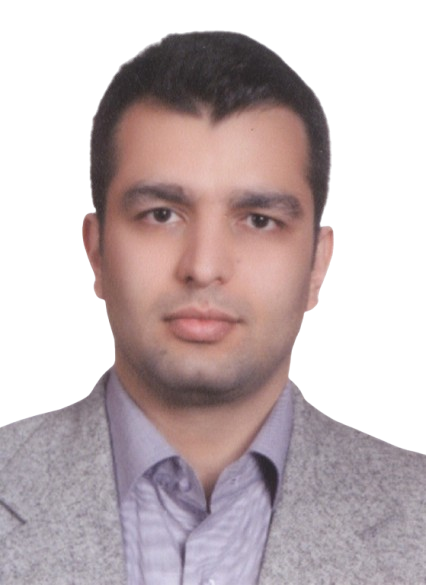}}]{Saeid Gorgin} (Senior Member, IEEE) received the B.S. degree in Computer Engineering from Azad University, South Tehran Branch, Tehran, Iran, in 2001, the M.S. degree in Computer Engineering from the Azad University, Science and Research Branch, Tehran, in 2004, and the Ph.D. degree in Computer System Architecture from Shahid Beheshti University, Tehran, in 2010.
He previously served as an Associate Professor of Computer Engineering in the Department of Electrical Engineering and Information Technology at the Iranian Research Organization for Science and Technology (IROST), Tehran. He was also a Research Professor with Sungkyunkwan University, South Korea. He is currently a Lecturer with the School of Physics, Engineering and Computer Science, University of Hertfordshire, Hatfield, U.K.
His research interests include applied machine learning, embedded AI, hardware accelerators, computer arithmetic, and VLSI design.
\end{IEEEbiography}

{\vskip -1\baselineskip plus -1fil}

\begin{IEEEbiography}[{\includegraphics[width=1in,height=1.25in,clip,keepaspectratio]{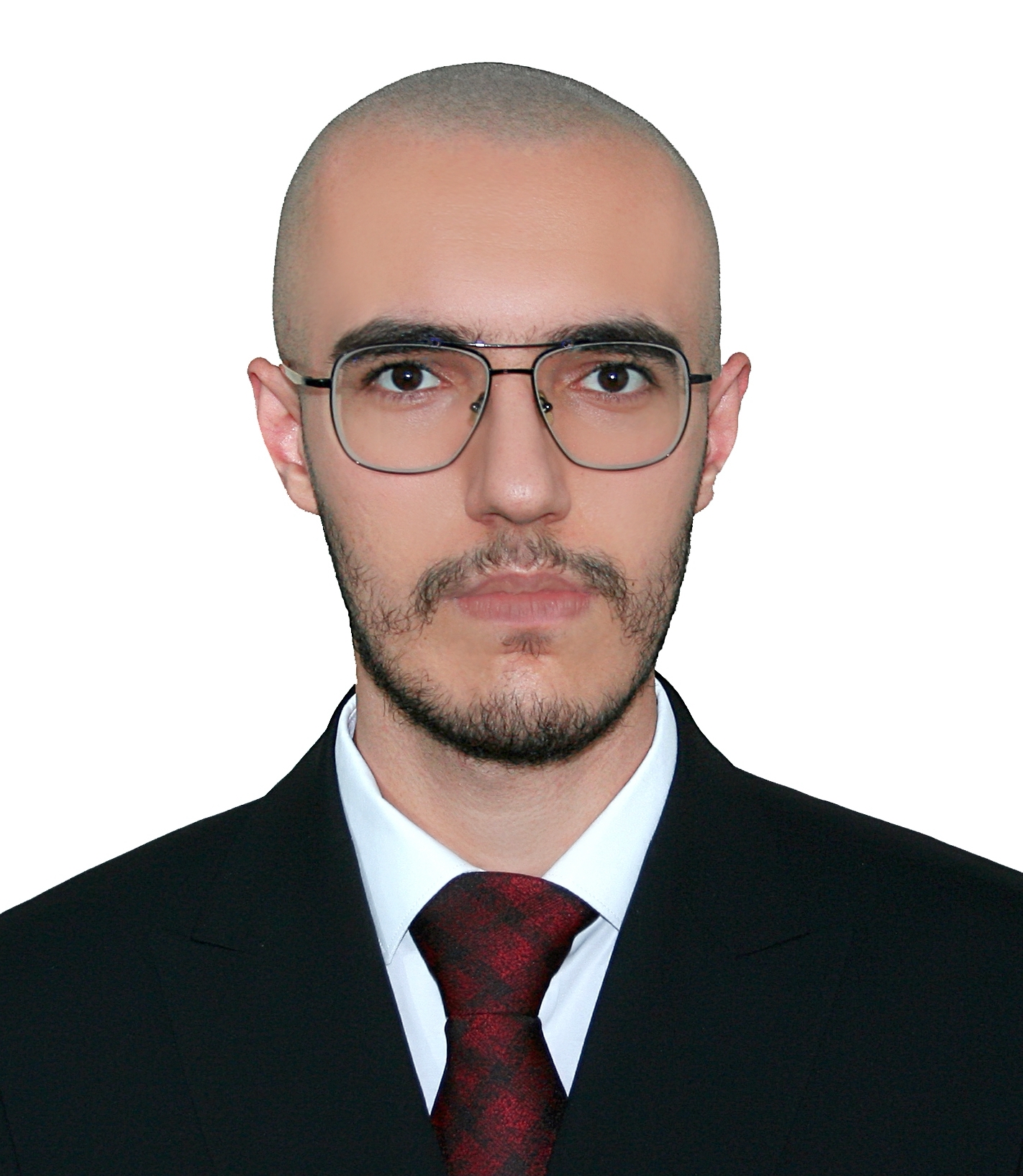}}]
{Amirhossein Sadr} (Student Member, IEEE)  
is currently an undergraduate student in Computer Engineering at Shahid Beheshti University, Tehran, Iran. 
He is also a Research Assistant at the High Performance Computing Center of the Institute for Research in Fundamental Sciences (IPM). His academic interests include unconventional computing, artificial intelligence, computer architecture, and hardware accelerators.
\end{IEEEbiography}

{\vskip -1\baselineskip plus -1fil}

\begin{IEEEbiography}[{\includegraphics[width=1in,height=1.25in,clip,keepaspectratio]{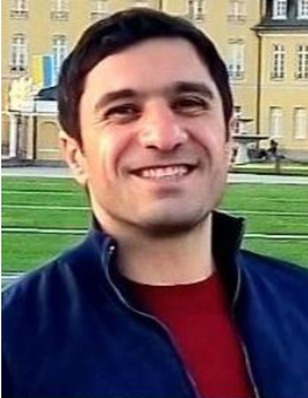}}]{Behzad Salami} is a Leading Researcher and head of the FPGATech Group at BSC, working on FPGA-based reconfigurable computing for pre-silicon emulation of RISC-V architectures, hardware acceleration, and system-level exploration of energy, dependability, and performance trade-offs. He received his Ph.D. in Computer Architecture with honors from UPC, where his work focused on low-power and fault-resilient accelerators for emerging workloads. He has published in leading venues like MICRO, DSN, DATE, IEEE MICRO, has contributed to multiple national and European research projects, and has received several research awards and recognitions for his works.
\end{IEEEbiography}

\begin{IEEEbiography}[{\includegraphics[width=1in,height=1.25in,clip,keepaspectratio]{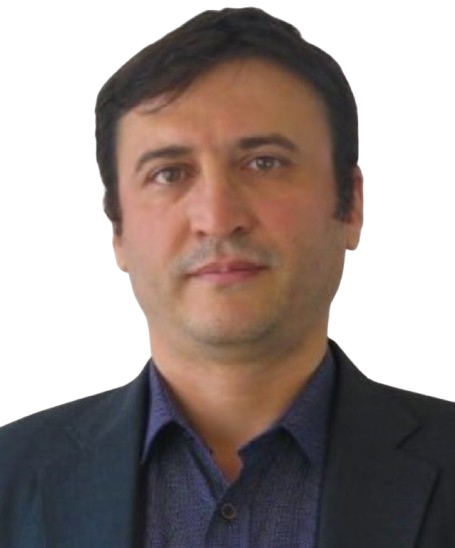}}]{Dara Rahmati} (Member, IEEE) received the B.Sc. and M.Sc. degrees in Computer Engineering from the University of Tehran, Iran, in 1998 and 2001, respectively, and the Ph.D. degree in Computer Engineering from Sharif University of Technology, Tehran, Iran.
He is currently an Assistant Professor in the Department of Computer Science and Engineering at Shahid Beheshti University, Tehran, Iran. His research interests include the design and performance evaluation of optimized networks-on-chip architectures, heterogeneous architectures, and hardware accelerators.
\end{IEEEbiography}

\vfill

\end{document}